\documentclass[floatfix,onecolumn,showpacs]{revtex4}
\pdfoutput=1
\usepackage{graphics,graphicx}
\usepackage{epsfig}
\usepackage{amsmath}
\usepackage{amstext}
\usepackage{amssymb}
\usepackage{dcolumn}

\newcommand{\beq}{\begin{eqnarray}}
\newcommand{\eeq}{\end{eqnarray}}

\begin{document}

\title{Emergence of $q$-statistical functions in a generalized binomial distribution with strong correlations}

\author{Guiomar Ruiz$^{1,2}$  and Constantino Tsallis$^{1,3}$}
\affiliation{$^1$Centro Brasileiro de Pesquisas Fisicas and \\National Institute of Science and Technology for Complex Systems,
Rua Xavier Sigaud 150, 22290-180 Rio de Janeiro-RJ, Brazil\\
$^2$ Departamento de Matem\'{a}tica Aplicada y Estad\'{\i}stica, Universidad Polit\'{e}cnica de Madrid, Pza. Cardenal Cisneros s/n, 28040 Madrid, Spain\\
$^3$Santa Fe Institute, 1399 Hyde Park Road, Santa Fe, NM 87501, USA}

\begin{abstract}
We study a symmetric generalization $\mathfrak{p}^{(N)}_k(\eta, \alpha)$ of the binomial distribution recently introduced by Bergeron et al, where $\eta \in [0,1]$ denotes the win probability, and $\alpha$ is a positive parameter.
This generalization is based on $q$-exponential generating functions ($e_{q^{gen}}^z \equiv [1+(1-q^{gen})z]^{1/(1-q^{gen})};\,e_{1}^z=e^z)$ where $q^{gen}=1+1/\alpha$.
The numerical calculation of the probability distribution function of the number of  wins $k$, related to the number of realizations $N$, strongly approaches a discrete $q^{disc}$-Gaussian distribution, for win-loss equiprobability (i.e., $\eta=1/2$) and all values of $\alpha$.
Asymptotic $N\to \infty$ distribution is in fact a $q^{att}$-Gaussian $e_{q^{att}}^{-\beta z^2}$,
where $q^{att}=1-2/(\alpha-2)$ and
$\beta=(2\alpha-4)$.
The behavior of the scaled quantity $k/N^\gamma$ is discussed as well. For $\gamma<1$, a large-deviation-like property showing a $q^{ldl}$-exponential decay is found, where $q^{ldl}=1+1/(\eta\alpha)$. For $\eta=1/2$, $q^{ldl}$ and $q^{att}$ are related through $1/(q^{ldl}-1)+1/(q^{att}-1)=1$, $\forall \alpha$. For $\gamma=1$, the  law of large numbers is violated, and we consistently study  the large-deviations with respect to the probability of the $N\to\infty$ limit distribution, yielding a power law, although not  exactly a $q^{LD}$-exponential decay. All $q$-statistical parameters which emerge are univocally defined by $(\eta, \alpha)$. Finally we discuss the analytical connection with the P\'{o}lya  urn problem.
\end{abstract}

 \maketitle

\section{Introduction}
\label{introduction}
Probability distributions that take correlations into account, can be in some cases constructed by deformation of mathematical independent laws \cite{Dodonov02,Gazeau09,Matos96,Kis01}. Along this line of approach, a variety of generalizations of the binomial distribution have been recently proposed \cite{Curado10, Bergeron12, Bergeron14}. The generalization consists in replacing the sequence of natural numbers that correspond to the variable of the original binomial distribution, by an arbitrary sequence of non-negative numbers. Consequently, their factorial and combinatorial numbers are redefined, in such a way that the simple powers of ``win'' (``loose'') probability must be replaced by characteristic polynomials whose degree is the number of wins (loose). These polynomials are obtained through generating functions that force the generalized distributions  to still satisfy the  conditions of normalization and non-negativeness. Resulting  probabilities $\mathfrak{p}^{(N)}_k$ can be symmetrical or asymmetrical,  and represent  the probability of having $k$ wins and $(N-k)$ losses, in a sequence of $N$ {\it  correlated} trials.

We shall be concerned with a particular set of {\it symmetrically} generalized binomial distributions, so as to preserve the win-loss symmetry, which is an essential prerequisite for the distribution to be used in the present analysis of strongly correlated systems and their entropic behavior. The  generating functions of this particular set are $q^{gen}$-exponentials ($e_{q^{gen}}^z \equiv [1+(1-q^{gen})z]^{1/(1-q^{gen})};\,e_{1}^z=e^z)$, where {\it gen} stands for {\it generating}, and $q^{gen}=1+1/\alpha >1$ ($\alpha>0$), $\alpha$ being a parameter to be soon defined. These generating functions are the  only ones, besides the ordinary binomial case (which corresponds $q^{gen}\to 1$), that yield probabilities which obey the Leibnitz triangle rule (see for instance \cite{Bergeron14,TsallisGellMannSato2005}). The family of the generated probabilities depends on  two parameters ($\eta$, $\alpha$), where $\eta$ is the ``win'' probability, $(1-\eta)$ is the ``loss" probability, and $\alpha$ characterizes the generating function. A variety of $q$-statistical functions  \cite{Tsallis88,Tsallis09} related to the probabilities $\mathfrak{p}^{(N)}_k(\eta, \alpha)$ emerges, all of them univocally defined by $(\eta, \alpha)$, and whose respective $q(\eta, \alpha)$ indices appear to obey  an algebra that reminds the underlying algebra in \cite{TsallisGellMannSato2005,Umarov10}.

In fact, such probabilities provide a probability distribution function of the scaled quantity $k/N$ ($k/N=0,1/N,\dots, 1$) that is  likely to achieve any of the complete set \cite{Rodriguez2014} of  bounded support $q^{disc}$-Gaussian $e_{q^{disc}}^{-\beta z^2}$ (where {\it disc} stands for {\it discrete}, and $q^{disc} <1$), where $\beta$ is a generalized inverse temperature ($\beta \in \mathbb{R}$). The $q$-Gaussian form corresponds to  strongly correlated random variables, and arises from the extremization of the nonadditive entropy $S_q=k_B(1-\sum_ip_i^q)/(q-1)$   ($q\in \mathbb{R}$,
 $S_{1}=S_{BG} \equiv -k\sum p_i \ln p_i$, where BG stands for Boltzmann-Gibbs) \cite{Tsallis88} under appropriate constraints \cite{Prato99, Umarov07}.

Since the proposal in \cite{Tsallis09}, several  statistical models which provide, in the $N\to\infty$ limit, $q$-Gaussian attractors have been constructed 
\cite{Hanel09,Rodriguez12,Ruseckas14}. Some of them exhibit  extensivity of  the Boltzmann-Gibbs entropy $S_{BG}$, and at least one of them exhibits extensivity of the $S_{q}$ entropy for $q \ne 1$ \cite{Ruseckas14}. In particular, the  $q^{gen}$-exponentially generated  probabilities $\mathfrak{p}^{(N)}_k(\eta, \alpha)$ exhibit an extensive $S_{BG}$ \cite{Evaldo2014}, and they appear to provide  $q^{att}$-Gaussian attractors (where $att$ stands for {\it attractor}).
An outstanding fact is that these  probabilities have been  rigorously deduced a priori, imposing a particular structure of the generating function (see also \cite{Carati2005, Carati2008}) under  non-negativeness and normalization probability conditions. In addition to these properties, these probabilities can be shown to correspond to the P\'{o}lya urn model  \cite{Polya}.

The paper is organized as follows. Section \ref{sec2} presents the symmetric generalized binomial distribution  introduced by H. Bergeron et al. \cite{Bergeron14}, based on the $q^{gen}$-exponential generating functions. Section \ref{sec3} is devoted to the characterization of the involved $q$-Gaussian distributions  (where $q$ refers to $q^{disc}$ for finite $N$, and to $q^{att}$ in the $N\to\infty$ limit)  in the  generalized probability distribution of the ratio (number of  wins)/(number of realizations). Section \ref{sec4} describes large-deviation-like properties of the distribution of  the scaled quantity $k/N^{\gamma}$ ($0<\gamma<1$). Section \ref{sec5} deals with the behavior of the large-deviation probabilities with regard to the $N\to \infty$ limit distribution, which violates the law of large numbers. We conclude in Section \ref{Conclusions}.

\section{The Symmetric Generalized Binomial Distribution}
\label{sec2}
In a sequence of $N\in \mathbb{N}$ independent trials with two possible outcomes, ``win'' and ``loss'', the probability of obtaining $k$ wins is given by the binomial distribution:
\begin{equation}p^{(N)}_k(\eta)=
\left(\begin{array}{c}N \\ k \end{array}\right)\eta^k(1-\eta)^{N-k}=
\frac{N!}{(N-k)! \, k!}\eta^k(1-\eta)^{N-k}
\end{equation}
where the parameter $\eta$ ($0\le \eta \le 1$) is the probability of having the outcome ``win'',  $(1-\eta)$ corresponding to the outcome ``loss''.
Therefore, the Bernoulli binomial distribution above preserves the symmetry win-loss.

Let us now consider  an strictly increasing infinite sequence of {\it nonnegative real} numbers $\chi=\{x_N\}_{N\in \mathbb{N}}$. With each sequence $\chi$ defined above, a Bernoulli-like distribution is constructed:
\begin{equation}
\mathfrak{p}^{(N)}_k(\eta)=
\frac{x_N!}{x_{N-k}! \, x_k!}q_k(\eta)  q_{N-k}(1-\eta)
\label{binlike}
\end{equation}
where the factorials are defined as $x_N!\equiv x_1x_2\dots x_N$, $x_0!\equiv 1$, $\eta$ is a running parameter on the interval $[0,1]$, and $q_k(\eta)$ are polynomials of degree $k$. Observe that the symmetry win-loss of binomial-like distribution (\ref{binlike}) is preserved, as invariance under $[k,\eta] \to [(N-k),(1-\eta)]$ is verified. That means that no bias can exist favoring either win or loss when $\eta=1/2$.

The polynomials  $q_k(\eta)$ are to be defined, is such a way that  quantities $\mathfrak{p}^{(N)}_k(\eta)$ represent the probabilities of having $k$ wins and $(N-k)$ losses in a sequence of {\it correlated $N$ trials}. Consequently, $\mathfrak{p}^{(N)}_k(\eta)$ must be constrained by
the normalization equation:
\begin{equation}\label{normal}\displaystyle\sum_{k=0}^N\mathfrak{p}^{(N)}_k(\eta)=1, \qquad  \forall N\in \mathbb{N},  \forall \eta \in [0,1],
\end{equation}
and  the non-negativeness condition:
\begin{equation}\label{nnegatv}\mathfrak{p}^{(N)}_k(\eta)\ge 0, \qquad \forall N,k\in \mathbb{N},  \forall \eta \in [0,1].\end{equation}

Different sets of polynomials can be associated with (\ref{normal}) and (\ref{nnegatv}). With this aim, some generating functions of polynomials can be considered. Let us make use of a $q^{gen}$-exponential generating function
$e_{q^{gen}}^z = [1+(1-q^{gen})z]^{1/(1-q^{gen})}$.
Such a generating function can be written as $\mathcal{N}(z)=\left(1-z/\alpha\right)^{-\alpha}$ ($\alpha >0$),  where the $q$-exponential parameter that characterizes  the generating function is $q^{gen}=1+1/\alpha>1$. The following  probability distributions are consequently obtained \cite{Bergeron14}:

\begin{equation}
\mathfrak{p}^{(N)}_k(\eta, \alpha)=
\left(\begin{array}{c}N\\k \end{array}\right)
\frac{\left(\eta \alpha \right)_k\left((1-\eta)\alpha\right)_{N-k}}{\left(\alpha\right)_N}
\label{distribevaldo}
\end{equation}
where $(a)_b \equiv a(a+1)(a+2)\dots (a+b-1)$
is the Pochhammer symbol, and $\alpha=1/(1-q^{gen})$. Observe that the $\alpha \to \infty$ limit recovers the ordinary binomial case ($q^{gen}\to 1$). The expectation value  and the variance of (\ref{distribevaldo}) are $\langle k\rangle_N(\eta)=\eta N$  and  $(\sigma_k)_N^2(\eta, \alpha)=N^2\eta(1-\eta)\displaystyle\frac{1+\alpha/N}{1+\alpha}$, respectively \cite{Bergeron14}.

In  the particular cases $\eta=1/2$ and $4\le \alpha=\dot{2}$ (i.e., $\alpha=4,6,8, \dots$), we can also use the following equivalent expression:
\begin{equation}
\mathfrak{p}^{(N)}_k(\eta=1/2, \alpha)=\left\{\begin{array}{ll}
\displaystyle\frac{(\alpha/2)_N}{(\alpha)_N}& (k= 0, N)\\
\left[\displaystyle\prod_{m=1}^{\alpha-1} \frac{\alpha-m}{N+\alpha-m}\right]\times\left[\displaystyle\prod_{j=1}^{\frac{\alpha}{2}-1} \left(\frac{k+j}{j}\right) \left(\frac{N-k+j}{j}\right) \right] & (k\ne 0, N) \,.
\end{array}\right. \label{improvenum}
\end{equation}
This expression is computationally very convenient.

\section{Emergence of  $q$-Gaussians}
\label{sec3}
\begin{figure}
\begin{center}
\includegraphics[width=8cm]{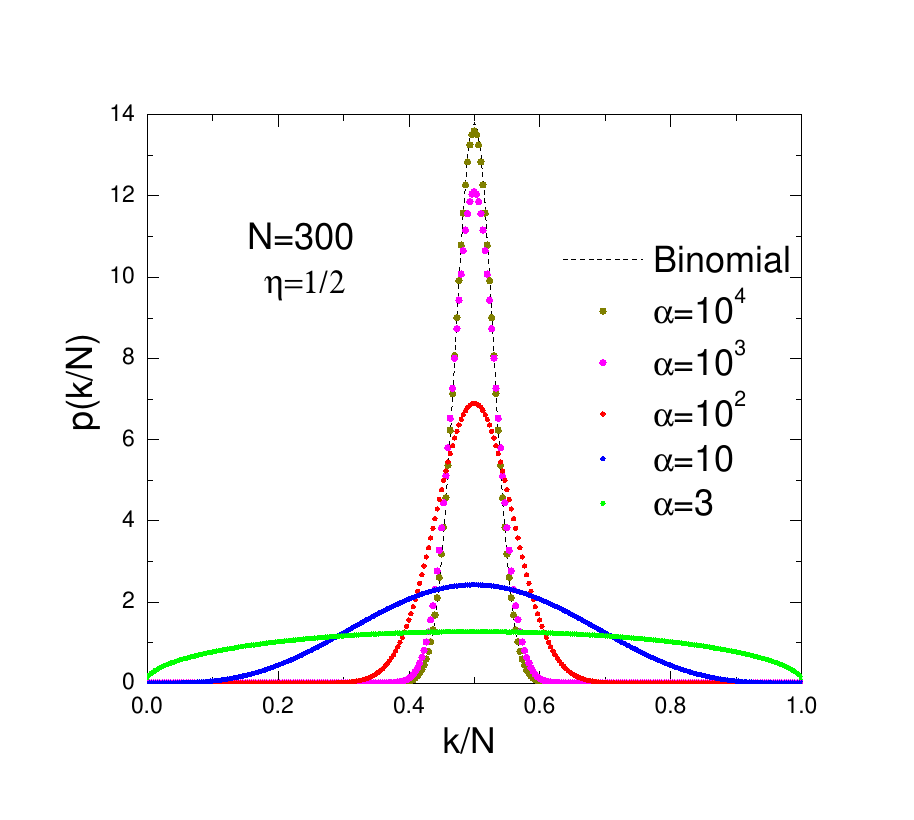}
\end{center}
\vspace{-0.5cm}
\caption{ Probability of having a ratio number of wins $k/N$ ($0 \le k/N \le 1$) in a sequence of $N=300$  correlated trials that follow the symmetric generalized binomial distribution. The parameters of the model are $\eta=1/2$, $\alpha=3,10,10^2,10^3, 10^4$. Observe that the limit distribution tends to a binomial distribution as $\alpha \to \infty$. If, in addition to this limit, we consider $N\to\infty$, the distribution shrinks onto a Dirac delta one. \label{fig1}}
\end{figure}

The histograms $p(k/N) \equiv N \mathfrak{p}^{(N)}_k(\eta, \alpha)$  ($0\le k/N \le 1$) are numerically obtained for fixed values of $N$, $\eta$ and $\alpha$. Fig.~\ref{fig1} shows that, for $\eta=1/2$  and $\alpha \to \infty$, $p(k/N)$  approaches the unbiased binomial distribution for all values of $N$.

\begin{figure}
\begin{center}
\includegraphics[width=7cm]{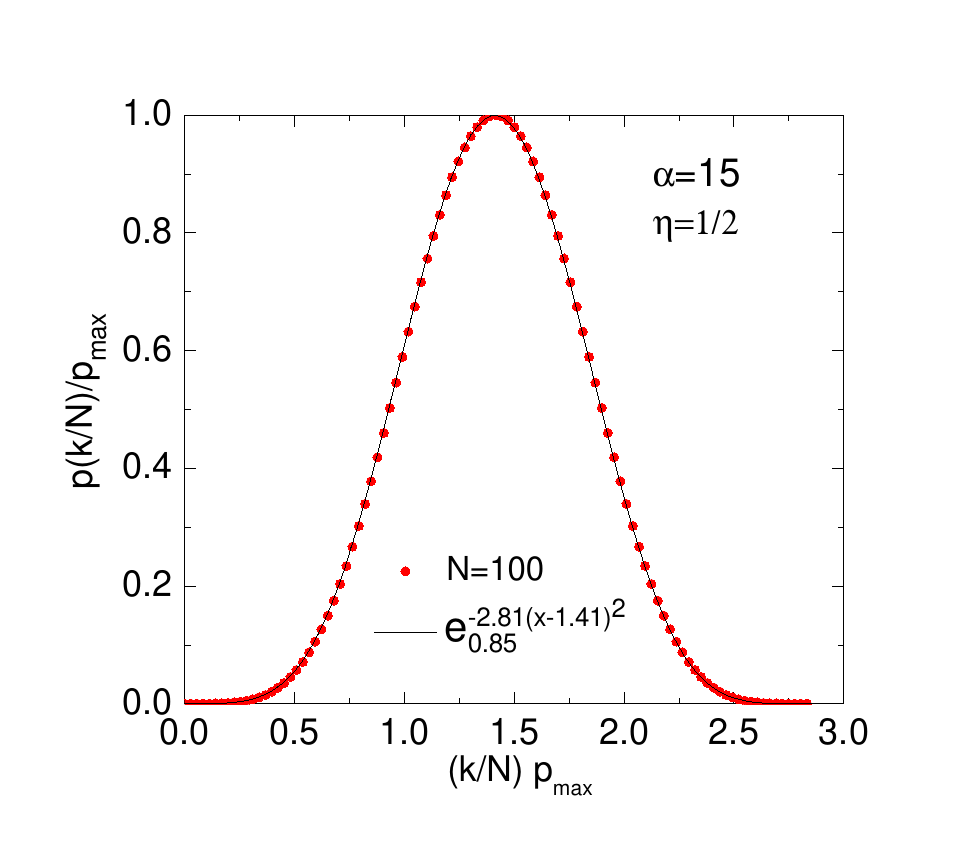}
\includegraphics[width=6.5cm]{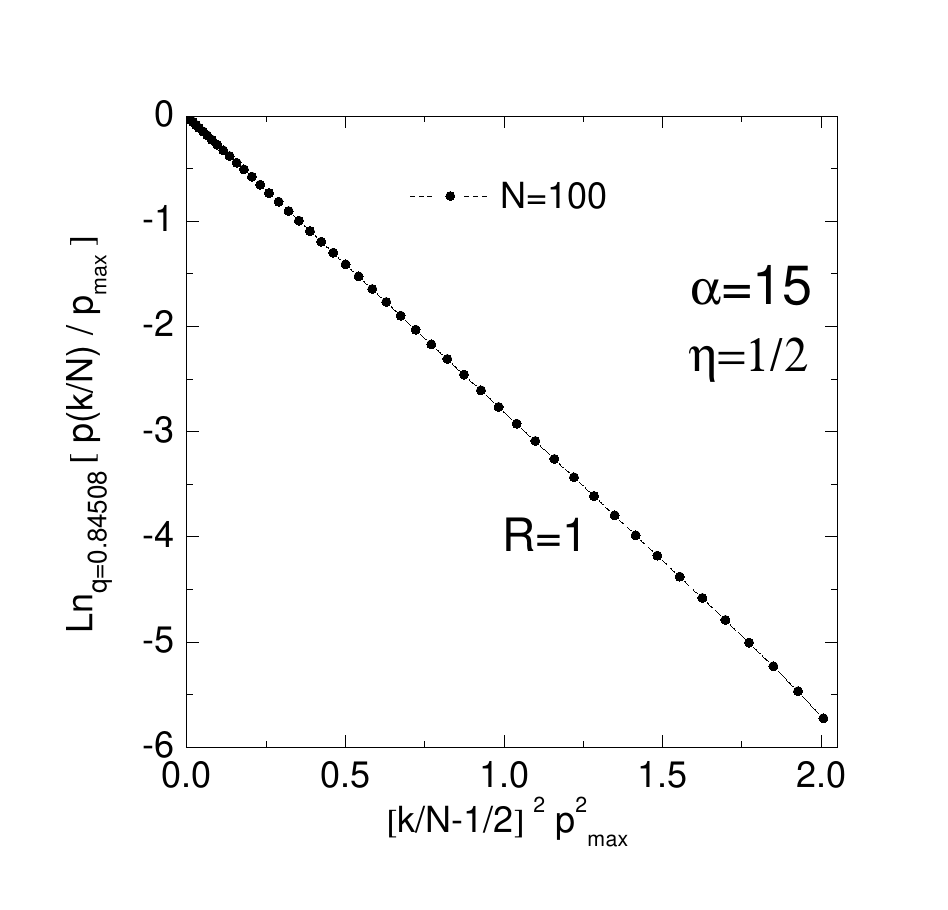}
\end{center}
\caption{ Normalized symmetric generalized binomial distribution for $\eta=1/2$, $N=100$ and $\alpha=15$. {\it Left panel}: The corresponding $q^{disc}$-Gaussian fitting function is superimposed. {\it Right panel}: The $q^{disc}$-logarithmic representation exhibits that indeed the discrete (i.e., $N<\infty$) distribution is extremely close to a $q$-Gaussian, the linear regression coefficient being $R=1$. \label{fig2}}
\end{figure}
Fig.~\ref{fig2} illustrates, for $\eta=1/2$ and a typical choice of $\alpha$ and $N$, the distribution $p(k/N)$ normalized to its maximum $P_{max}$. Our numerical results strongly suggest that $p(k/N)$ can be fitted by a $q^{disc}$-Gaussian distribution with  $q^{disc}=0.84508$. Other  values of $\alpha$ and $N$ have been  studied as well and, in all cases, numerical results strongly suggest  $q^{disc}$-Gaussian distributions (with $q^{disc}<1$), i.e.,
\begin{equation}
p(k/N)/p_{max}\simeq  e_{q^{disc}}^{-\beta (\frac{k}{N}-\frac{1}{2})^2}=\left[1-\beta (1-q^{disc})\Bigl(\frac{k}{N}-\frac{1}{2} \Bigr)^2\right]_+^{\frac{1}{1-q^{disc}}}
\label{qgaussian}
\end{equation}
where $[x]_+=x$ if $x>0$, and $[x]_+=0$ otherwise.
Table~\ref{table1} shows the values of $q^{disc}$  for typical values of $\alpha$ and $N$ ($\eta=1/2$ in all cases).
\begin{table}{}
\vspace{0.5cm}
{\small
\hspace{-0.4cm}
\begin{tabular}
{c|c|c|c|c|c|c|c|c|c}
 & $N=50$ & $N=60$& $N=70$ & $N=80$ & $N=100$& $N=200$ & $N=500$ &  $N=1000$
 &
 $N\to \infty$\\
\hline
$\alpha=3$ & $-0.98990$  & $-0.99170$& $-0.99306$ & $-0.99410$ &$-0.99524$&$-0.99800$  &  $-0.99930$& $-0.99970$&
$-1$
\\
\hline
$\alpha=5$ & $0.33117$  & $0.33185$& $0.33214$ & $0.33233$ &$0.33264$&$0.33311$  &  $0.33324$& $0.33328$
&$1/3$
\\
\hline
 $\alpha=15$  & $0.84235$ & $0.84350$ & $0.84412$ & $0.84456$    & $0.84508$ &$0.84587$&$0.84611$& $0.84614$ 
 &$11/13$ \\
\hline
 $\alpha=25$   & $0.90803$ &$0.90939$ & $0.91027
$ & $0.91084
$  & $0.91154
$& $0.91263
$ & $0.91297
$&$0.91303
$ 
&$21/23$\\
\hline
 $\alpha=50$  & $0.95127$ & $0.95298$&  $0.95413$& $0.95494$  & $0.95598$ &$0.95763$&$0.95821$&  $0.95830$ 
 &$23/24$\\
\hline
 $\alpha=100$ & $0.97043$
&$0.97244$  & $0.97382$ &   $0.97483$   & $0.97616$ & $0.97845$&$0.97936$& $0.97953
$
&$48/49$\\
\hline
 $\alpha=500$  & $0.98375$  & $0.98597$ & $0.98756$ & $0.98874$ &  $0.99039$ &0.99359& $0.99531$& $0.99576$
 &$248/249$\\
 \hline
 $\alpha=1000$  & $0.98521$ & $0.98745$ & $0.98904$ & $0.99025$  &  $0.99190$ &$0.99521$&$0.99710$ & $0.99766$
 &$498/499$\\
\hline
 $\alpha\to \infty$ & $1$
&$1$  & $1$ &   $1$   & $1$ & $1$&$1$& $1
$&$1$\\
 \hline
\end{tabular}} 
\caption{Numerical $q^{disc}$ parameter of the $q^{disc}$-Gaussian distribution strongly suggested by  $p(k/N)=N \mathfrak{p}^{(N)}_k$ for typical values of $\alpha$ and  $N$. The last column corresponds to the respective values of $\lim_{N\to \infty} q^{disc}$ (see text). }\label{table1}
\end{table}
For a fixed value of $N$, $q^{disc}$ is a monotonous function of $\alpha$.
Similarly, for a fixed value of $\alpha$,  $q^{disc}$ is  a monotonous function of $N$.

\begin{figure}
\begin{center}
\includegraphics[width=7cm]{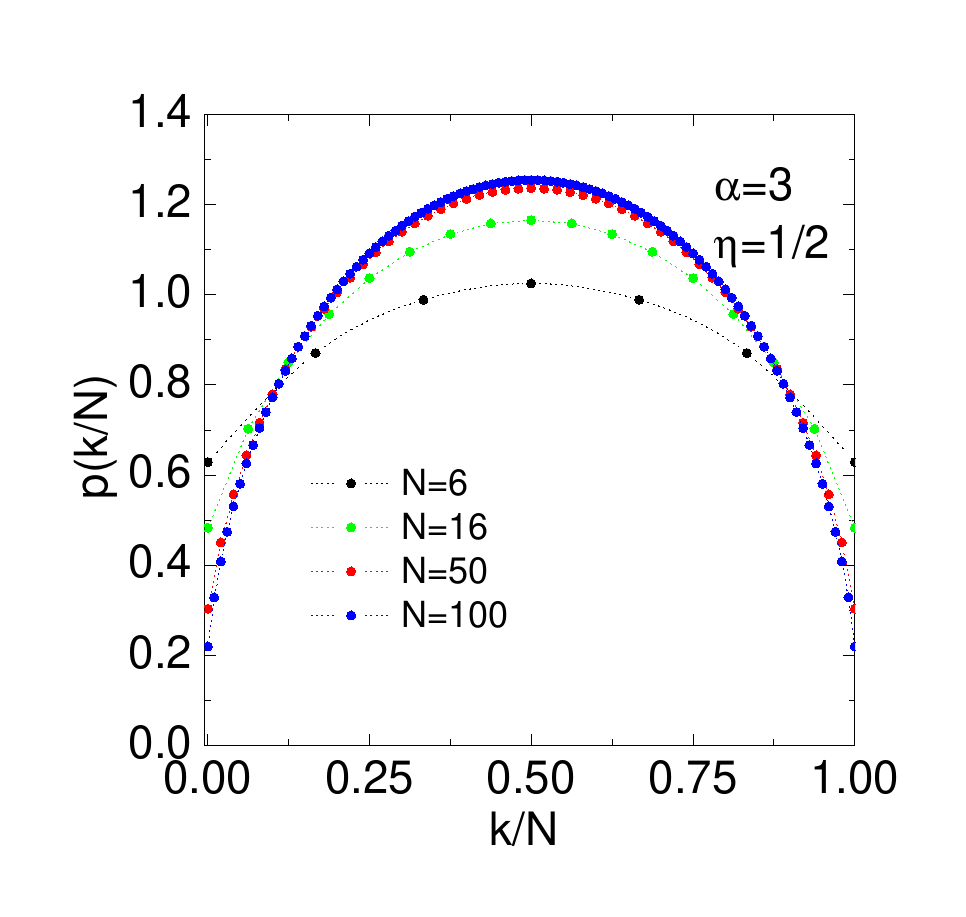}
\includegraphics[width=7cm]{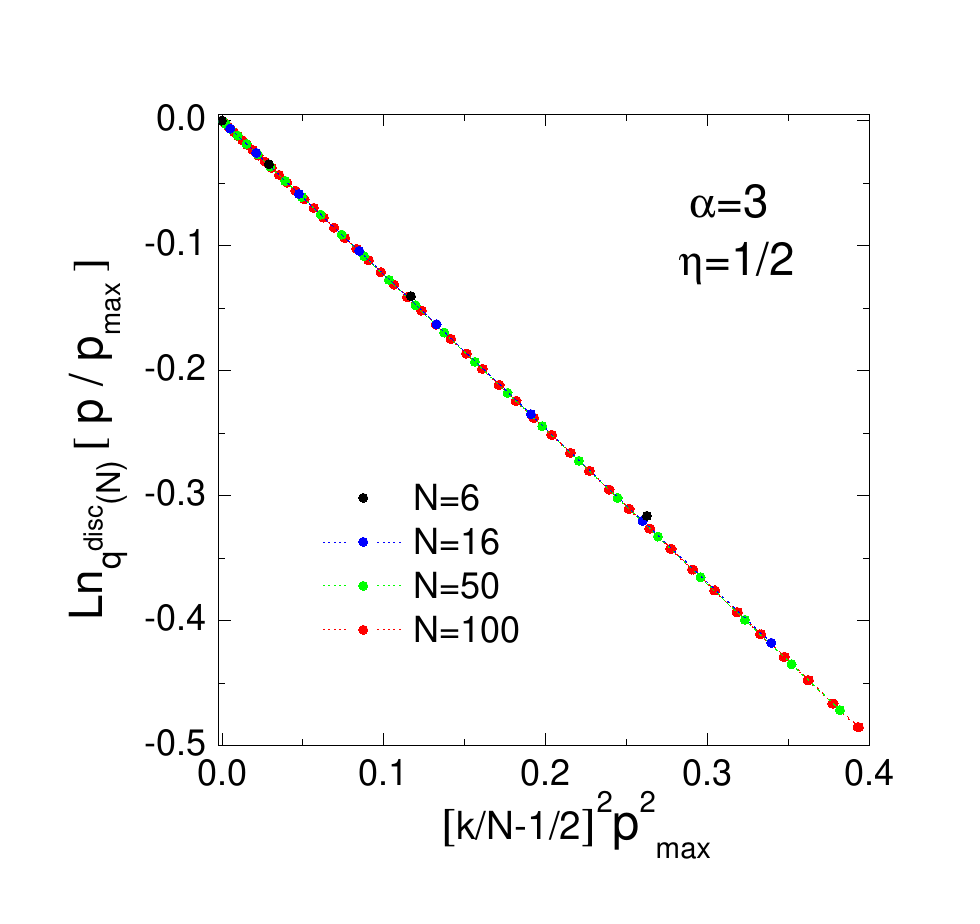}\\ \vspace{-1cm}
\includegraphics[width=7cm]{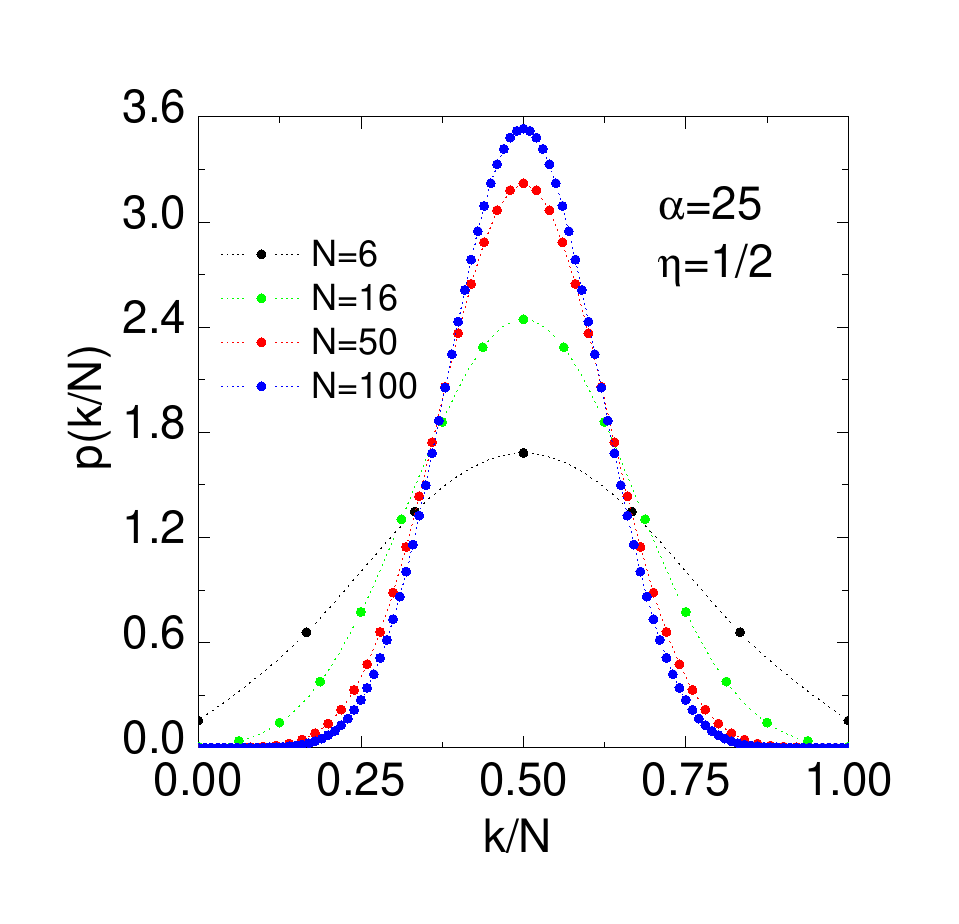}
\includegraphics[width=7cm]{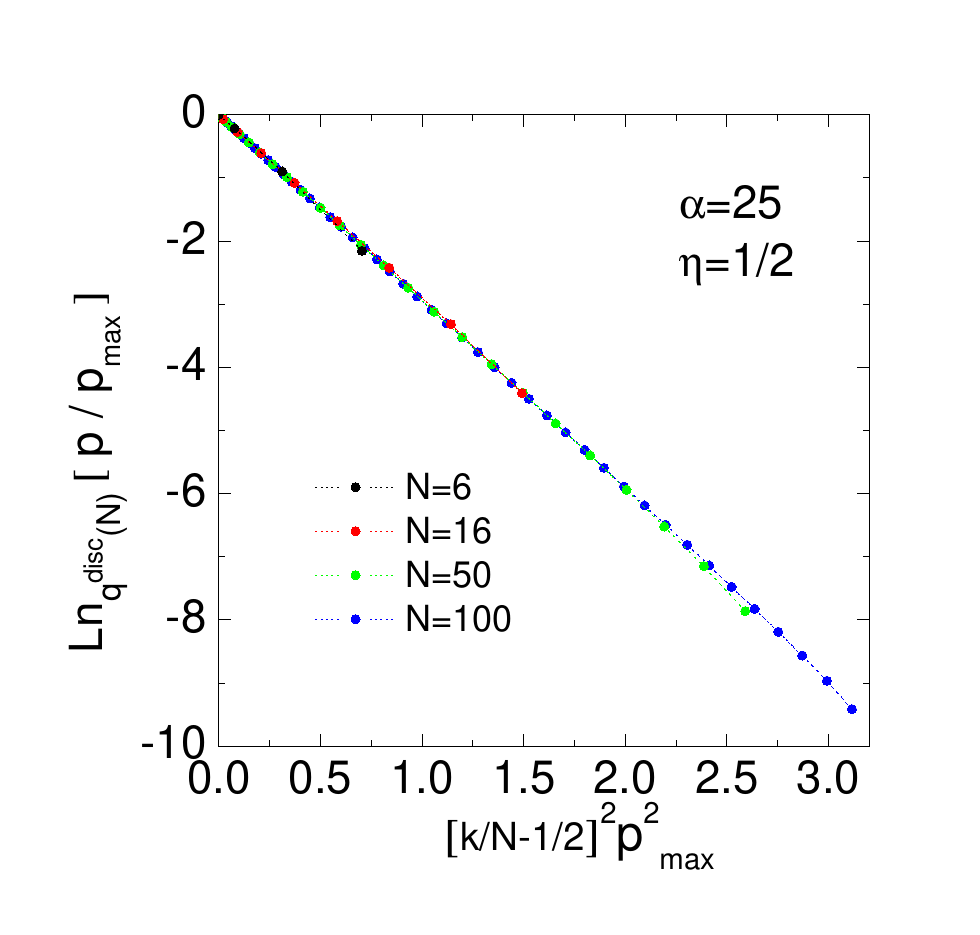}\vspace{-0.5cm}
\end{center}
\caption{{\it Left panels}: Probability  distribution    $p(k/N)=N\mathfrak{p}^{(N)}_k$ ($0\le k/N\le 1$) for typical values of $\alpha$ and $N$. {\it Right panels}:  The corresponding $q^{disc}(N)$-logarithmic representations show that the discrete distributions are very close to $q$-Gaussians, except for the last point of the tail. Similar results are obtained for other parameter values.  We notice that for $\alpha >4$ ($\alpha <4$), which corresponds to $q^{att} >0$ ($q^{att} <0$), the terminal derivative in the linear-linear representation  vanishes (diverges). This derivative is finite for $\alpha=4$, which corresponds to $q^{att}=0$.  \label{fig3}}
\end{figure}

Increasing the value of $N$, a sequence of $p(k/N)$ distributions is obtained (see Fig.~\ref{fig3}). Their corresponding $q^{disc}(N)$-logarithmic representation show that  $q^{disc}$-Gaussian distributions fit very well the data.

\begin{figure}
\begin{center}
\includegraphics[width=7cm]{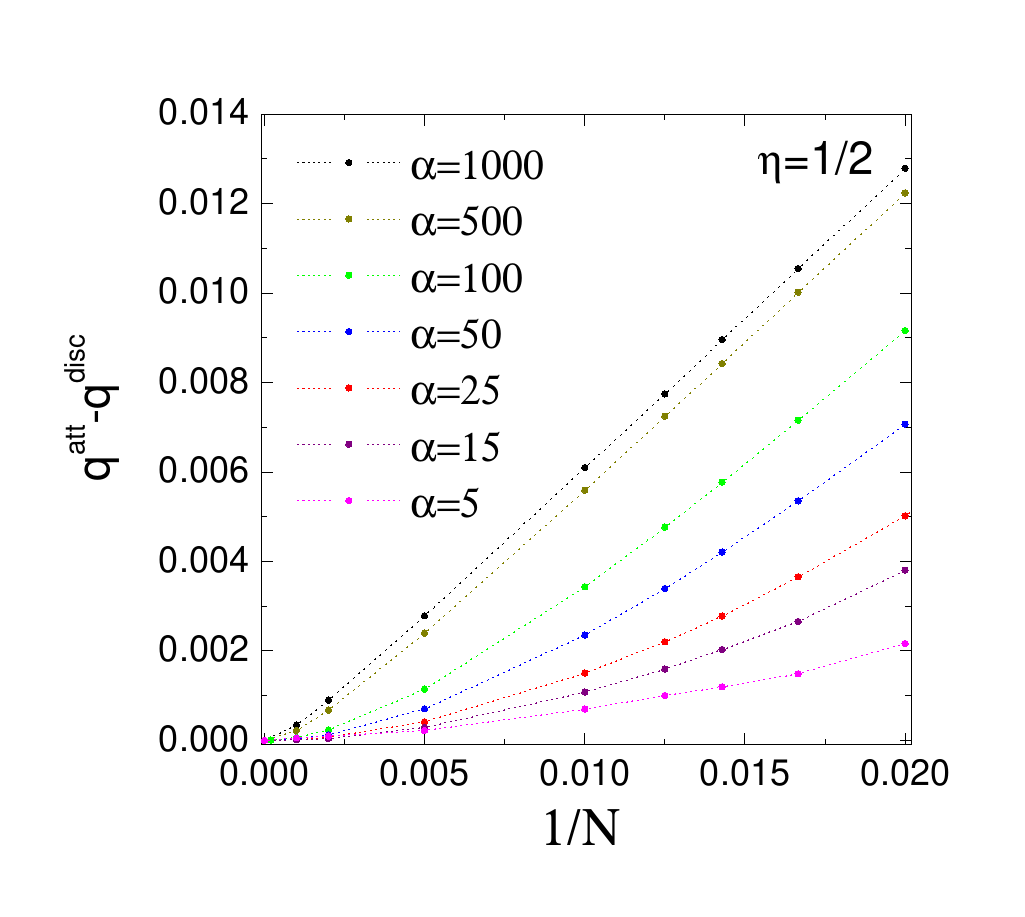}\vspace{-0.5cm}
\end{center}
\caption{Convergence  of the $q^{disc}$ index to the  $q^{att}$ index. \label{deltaqeff}}
\end{figure}

In the limit  $N\to \infty $,  we define
\begin{equation}
q^{att}(\alpha) \equiv \displaystyle\lim_{N \to \infty}q^{disc}(\alpha, N),
\end{equation}
and the corresponding $q^{att}$-Gaussian ($N\to \infty$) limit distributions are characterized by  $(q^{att}, \beta)$. This result that can in fact be   rigorously proved \cite{Evaldo2014}. Fig.~\ref{deltaqeff} shows the $N\to \infty$ convergence of the index $q^{disc}$ to the  index $q^{att}$.

\begin{figure}
\begin{center}
\includegraphics[width=7cm]{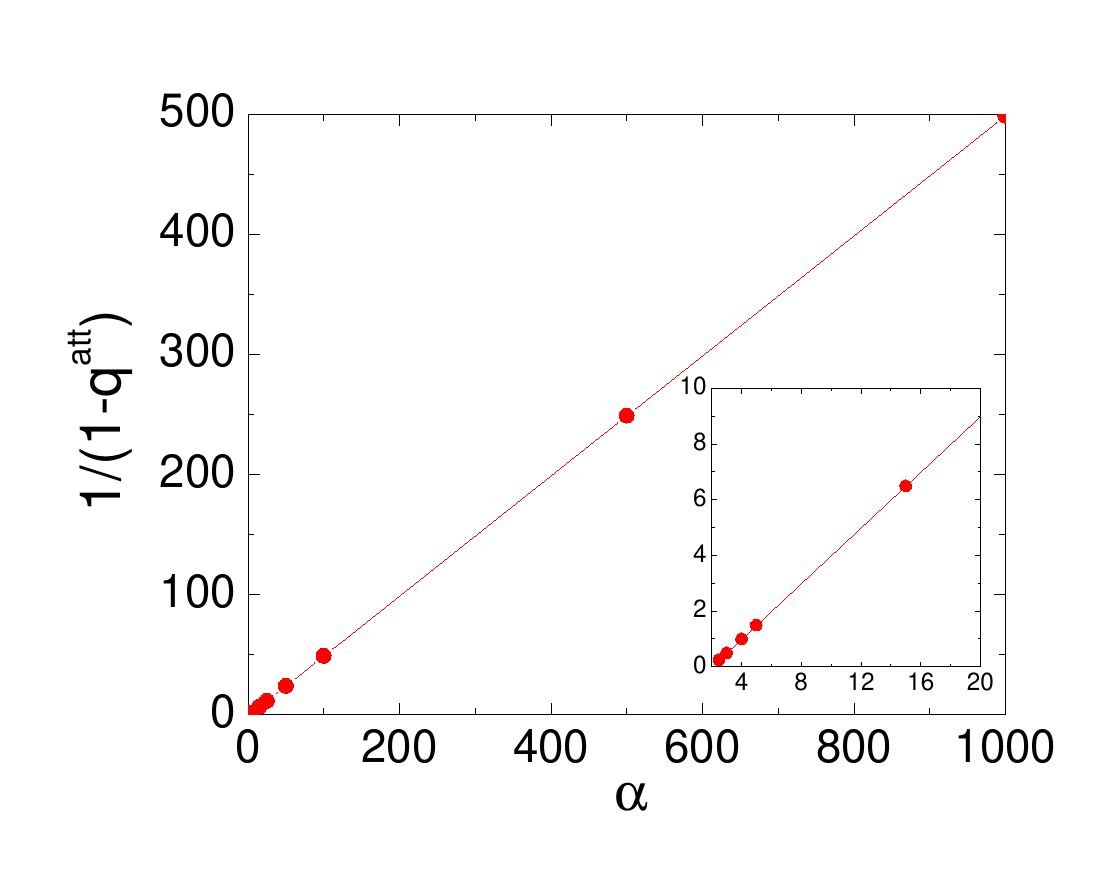}\vspace{-0.5cm}
\end{center}
\caption{The $\alpha$-dependence of $q^{att}$.\label{qalphaAtt}}
\end{figure}

The value of $q^{att}$ depends only on the parameter  $\alpha$, and  some values of the ($q^{att}, \alpha$) pair are shown in Table~\ref{table2}. We heuristically conclude (see Fig.~\ref{qalphaAtt}) that
\begin{equation}
q^{att}(\alpha)=1-\frac{2}{\alpha-2} \,.
\label{qalfa}
\end{equation}
\vspace{0.3cm}
\begin{table}\begin{center}
\begin{tabular}
{c|c|c|c|c|c|c|c|c|c|c|c|c}
$\alpha$  & $2$ & $2.1$& $2.5$ & $3$ & $4$& $5$ & $15$ &  $25$ &  $50$ & $100$&  $500$& $1000$
\\ \hline$q^{att}$&$-\infty$& $-19$&$-3$&$-1$& $0$&$1/3$&$11/13$&$21/23$&$23/24$ &$48/49$& $248/249$ &$498/499$\\
\end{tabular}
\end{center}
\vspace{-0.3cm}
\caption{Parameter $q^{att}(\alpha)= \displaystyle\lim_{N \to \infty}q^{disc}(\alpha, N)$ that characterizes the  $N\to \infty$ limit $q^{att}$-Gaussian distribution.}\label{table2}
\end{table}

\begin{figure}
\begin{center}\hspace{-0.5cm}
\includegraphics[width=6cm]{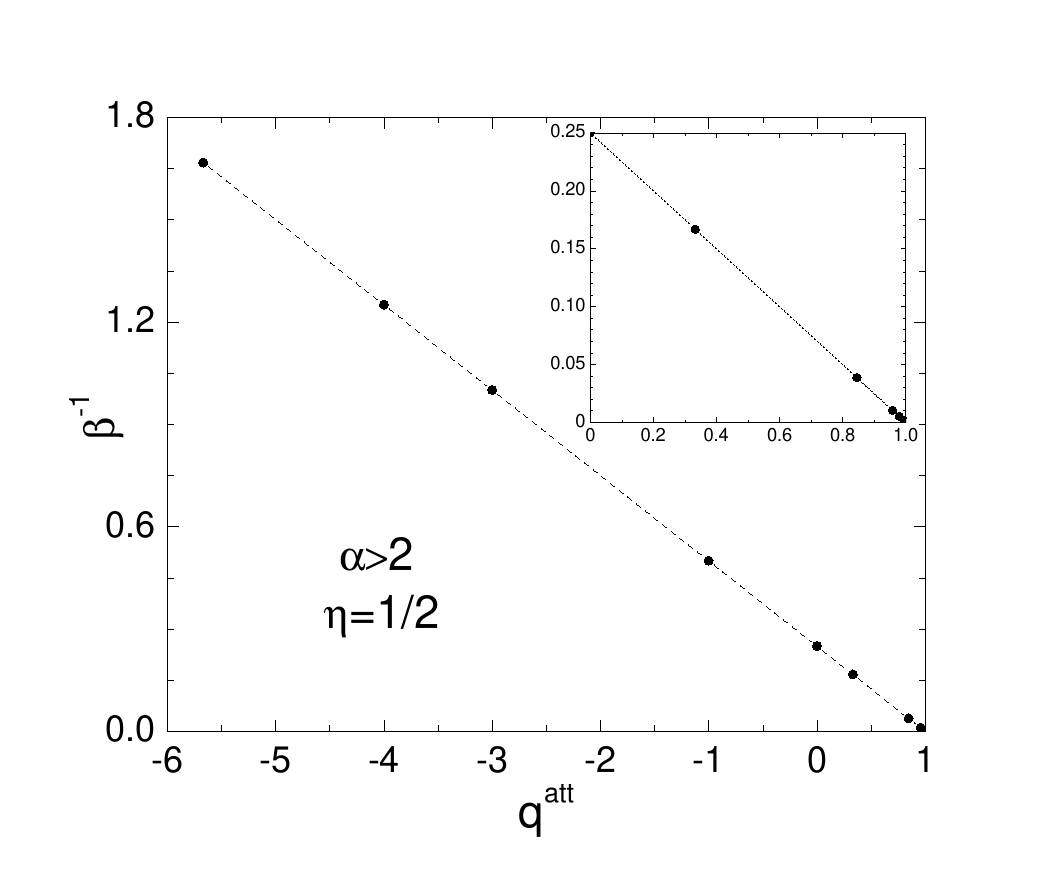}\hspace{-0.8cm}
\includegraphics[width=6cm]{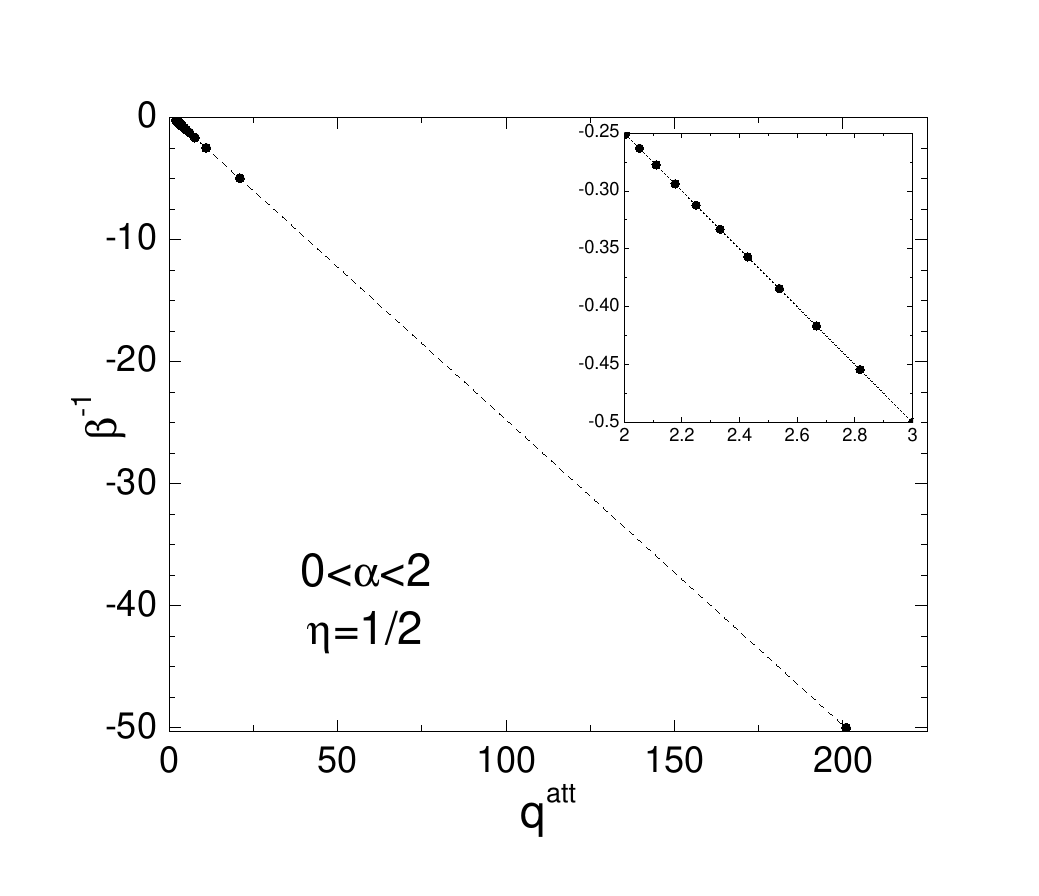}\hspace{-0.8cm}
\includegraphics[width=6cm]{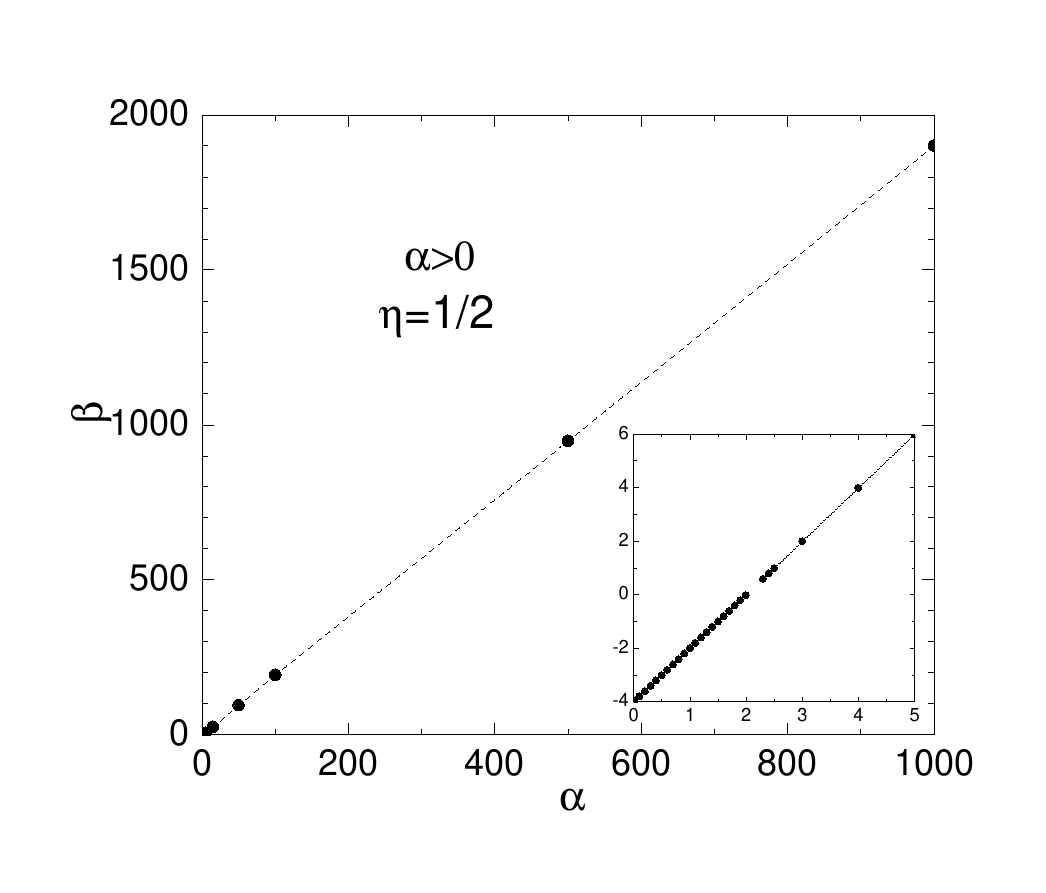}\vspace{-0.5cm}
\end{center}
\caption{Non-normalized histogram demonstrates simple $\beta(q^{att})$ and $\beta(\alpha)$  dependence for the  $q^{att}$-Gaussian limit distribution. Left panel:  $\alpha>2$ provides positive generalized temperature of  the  $q^{att}$-Gaussian. Central panel:  $0<\alpha<2$ provides negative generalized temperature of  the  $q^{att}$-Gaussian. Right panel: $\alpha$-dependence of the inverse temperature $\beta$ of  $q^{att}$-Gaussian. \label{BetaAtt}}
\end{figure}

Fig.~\ref{BetaAtt} shows the generalized temperature $\beta^{-1}$ of  non-normalized $q^{att}$-Gaussian distribution, as they are numerically obtained from $\alpha$. The $\alpha>2$ values correspond to positive values of $\beta$, and the $0<\alpha<2$ values correspond to negative values of $\beta$. In both cases, the following equation is satisfied:
\begin{equation}
\beta^{-1}=\frac{1-q^{att}}{4}\,.
\label{bq}
\end{equation}
This corresponds to the cut-offs of the distributions.
We can infer, from (\ref{qalfa}) and (\ref{bq}),  the following simple relation:
 \begin{equation}
\beta(\alpha)=2(\alpha-2) \,\,\,\, (\alpha >0)\,,
\label{balfa}
\end{equation}
that can also be checked in  Fig.~\ref{BetaAtt} (right panel).

\begin{figure}
\begin{center}
\includegraphics[width=7.1cm]{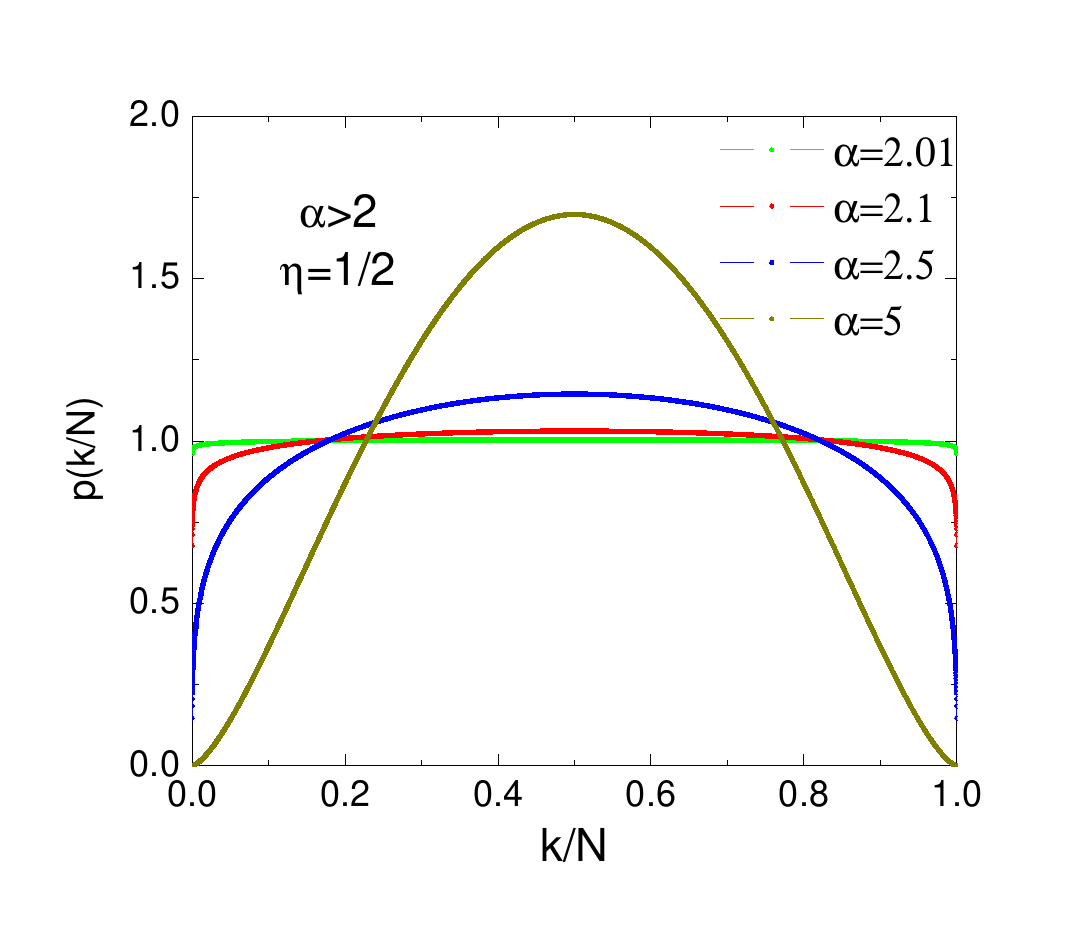}\hspace{-1cm}
\includegraphics[width=7cm]{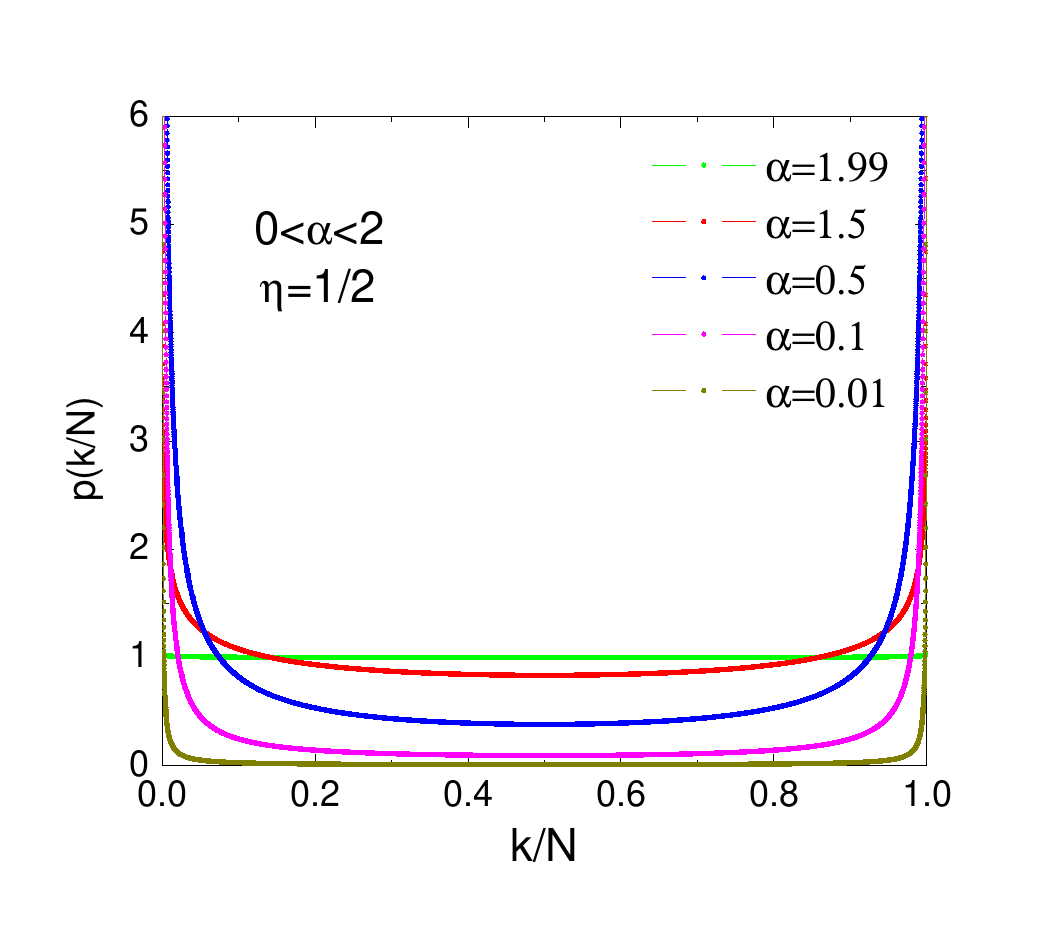}
\end{center}\vspace{-0.5cm}
\caption{{\it Left panel}: The $\alpha>2$ ($N\to \infty$) limit distributions, are concave with bounded and compact support. {\it Right panel}: The $0<\alpha<2$ ($N\to \infty$) limit distributions, are convex with bounded but non compact support. The limiting case $\alpha=2$ corresponds to the uniform distribution. \label{HTempNegtva1}}
\end{figure}

Summarizing, Fig.~\ref{HTempNegtva1} shows that  $\alpha> 2$ provides bell-shaped and compact support   
$q^{att}$-Gaussian distributions ($q^{att}<1$, $\beta>0$),   and $\alpha\in (0,2)$ provides convex and bounded but non compact support   $q^{att}$-Gaussian distributions ($q^{att}>2$, $\beta<0$). In the  $\alpha \to 2$ limit, an uniform  distribution is obtained.  This distribution is the limit of a $q^{att}$-Gaussian distribution whose $\displaystyle\lim_{\alpha\to 2^+}q^{att}(\alpha)=-\infty$ with $\beta>0$, as well as  $\displaystyle\lim_{\alpha\to 2^-}q^{att}(\alpha)=+\infty$ with $\beta<0$. In the $\alpha\to 0$ limit a double peaked delta distribution emerges,  i.e., the distribution is the limit of  a $q^{att}(\alpha)$-Gaussian distribution whose $\displaystyle\lim_{\alpha\to 0}q^{att}(\alpha)=2$ with $\beta<0$.
This description illustrates the diagram presented in \cite{Rodriguez2014}, where negative generalized temperatures $\beta^{-1}$ of $q$-Gaussian distributions are also considered.

The indexes  $q^{att}$ and $q^{gen}$ are related by
\begin{equation}
\frac{1}{q^{gen}-1}-\frac{2}{q^{att}-1}=2,
\end{equation}
equation that reminds the equations of the algebra indicated in \cite{Umarov10}.

Making use of the  expression of a normalized $q$-Gaussian \cite{Prato99}, it can be consequently stated that the generalized  distribution $p(k/N)$ defined in Eq.~\eqref{distribevaldo} corresponds, for win-loss equiprobability (i.e., $\eta=1/2$), to the following $q^{att}$-Gaussian  limit distribution:
\begin{equation}
p_{\infty}(x; \alpha)
\equiv\displaystyle\lim_{N\to \infty} p\left(\frac{k}{N}; \alpha\right)=\frac{2 \, \Gamma\left(\frac{5-3q^{att}}{2(1-q^{att})}\right)}{\sqrt{\pi} \, \Gamma\left(\frac{2-q^{att}}{1-q^{att}}\right)}\left[1-4 \left(x-\frac{1}{2}\right)^2\right]_+^{\frac{1}{1-q^{att}}}
\label{qgausianeq}
\end{equation}
where $\alpha>0$, $q^{att}=1-2/(\alpha-2)$, and $0\le x\le 1$.

\section{Large-deviation-like properties}
\label{sec4}
Let us now consider that  each of the $N$ single variables  takes values  $1$ or $0$ (i.e.,``win'' or ``loose''). Consequently, the value of $k$ corresponds to the sum of all $N$ binary random variables, and, after centering and re-scaling $k$, the attractors that emerge correspond to the abscissa currently associated with central limit theorems. In the case of the unbiased symmetric generalized probability distribution (i.e., $\eta=1/2$) defined in Eq.~\eqref{distribevaldo}, the  abscissa measured from its central value scales as  $N^{\gamma}$ with $\gamma =1$, and emerging attractors of $p(k/N)$ are the  $q^{att}$-Gaussians defined in Eq.~\eqref{qgausianeq}.
This fact precludes the vanishing limit of the  probability  $P$ of a deviation of $k/N$  from its central value $k/N=1/2$, i.e.,
\begin{equation}
\lim_{N\to \infty}P(N; |k/N-1/2| \ge \epsilon)
\ne 0 \, \, \,  (\forall \epsilon >0),\end{equation}
where $\epsilon$ is the minimum deviation of  $k/N$ with respect to  central value $k/N$, and $P$ can be evaluated  adding up the weight of the possible values of $k$ which do not fall inside $(N/2-\epsilon, N/2+\epsilon )$. Taking into account the symmetry of the distribution, and denoting
$x\equiv 1/2-\epsilon$, we can write
\begin{equation}
P(N; |k/N-1/2| \ge \epsilon)
=2P(N; k/N\le x)=
2\sum_{k=0}^{\lfloor N x \rfloor}\mathfrak{p}^{(N)}_k
\end{equation}
and conclude that, $\forall \alpha>0$,
\begin{equation}
\lim_{N\to \infty} P(k/N\le x;\alpha)=\lim_{N\to \infty} \sum_{k=0}^{\lfloor N x \rfloor}\mathfrak{p}^{(N)}_k( \alpha)
= \int_{0\le x \le \frac{k}{N}} p_{\infty}(x; \alpha) dx\ne 0.
\label{Largenumber}
\end{equation}
where $\lfloor N x \rfloor$ is the largest integer number that  $\lfloor N x \rfloor\le Nx$, and $p_{\infty}(x; \alpha)$ is defined in Eq.~\eqref{qgausianeq}. Eq.~\eqref{Largenumber} states that these  correlated models do {\it not} satisfy the classical version of the weak law of large numbers.

Due to this fact,
let us consider instead the scaled quantity $k/N^\gamma$,  $\gamma \ne 1$, in analogy with the anomalous diffusion coefficient introduced in nonlinear Fokker-Plank equations  \cite{FPlank}. More precisely, the case $\gamma \ne 1$ is similar to anomalous diffusion,  where the square space is nonlinear with time.

\begin{figure}
\begin{center}
\includegraphics[width=15cm]{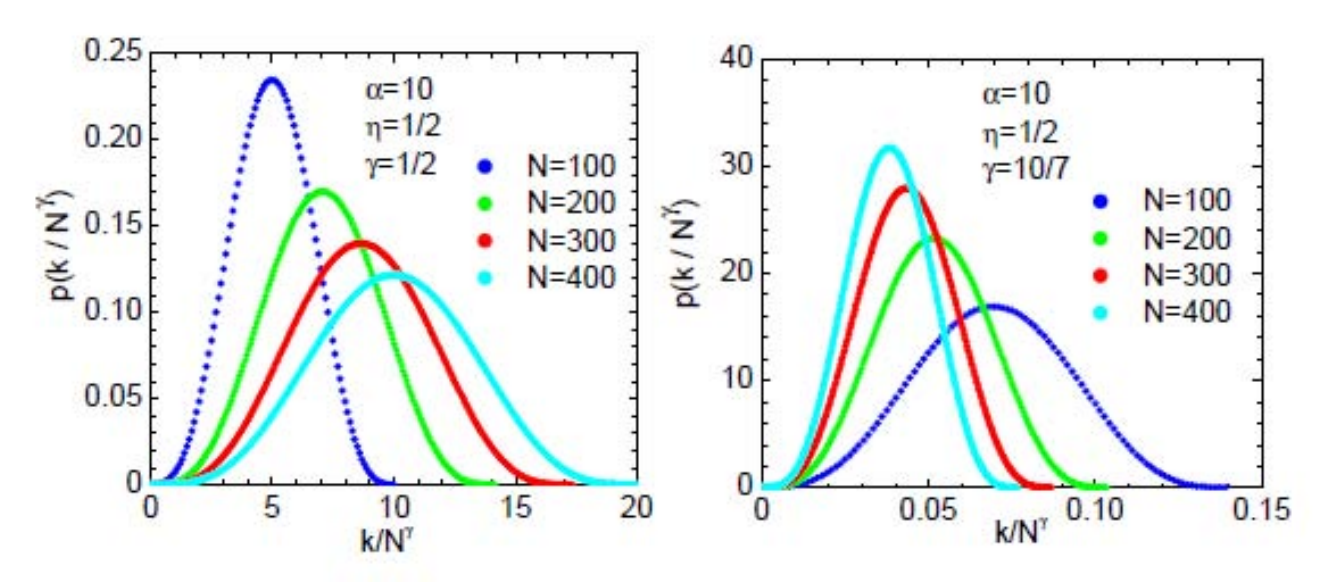}
\end{center}
\vspace{-0.5cm}
\caption{Non-centered histograms  of $k/N^\gamma$, $p(k/N^\gamma)=N^\gamma \mathfrak{p}^{(N)}_k$, for $\eta=1/2$, $\alpha=10$, and typical sequences of $N$. {\it Left panel:} Scaling factor $\gamma<1$ makes   $\lim_{N\to \infty}p(k/N^\gamma)=0$, $ \forall (k/N^\gamma)\in \mathbb{R}^+$. {\it Right panel:} Scaling factor $\gamma>1$ makes $\lim_{N\to \infty}p(k/N^\gamma)= \delta(0)$.
\label{Histgama}}
\end{figure}

\begin{figure}
\begin{center} \vspace{-0.5cm}
\includegraphics[width=5.7cm]{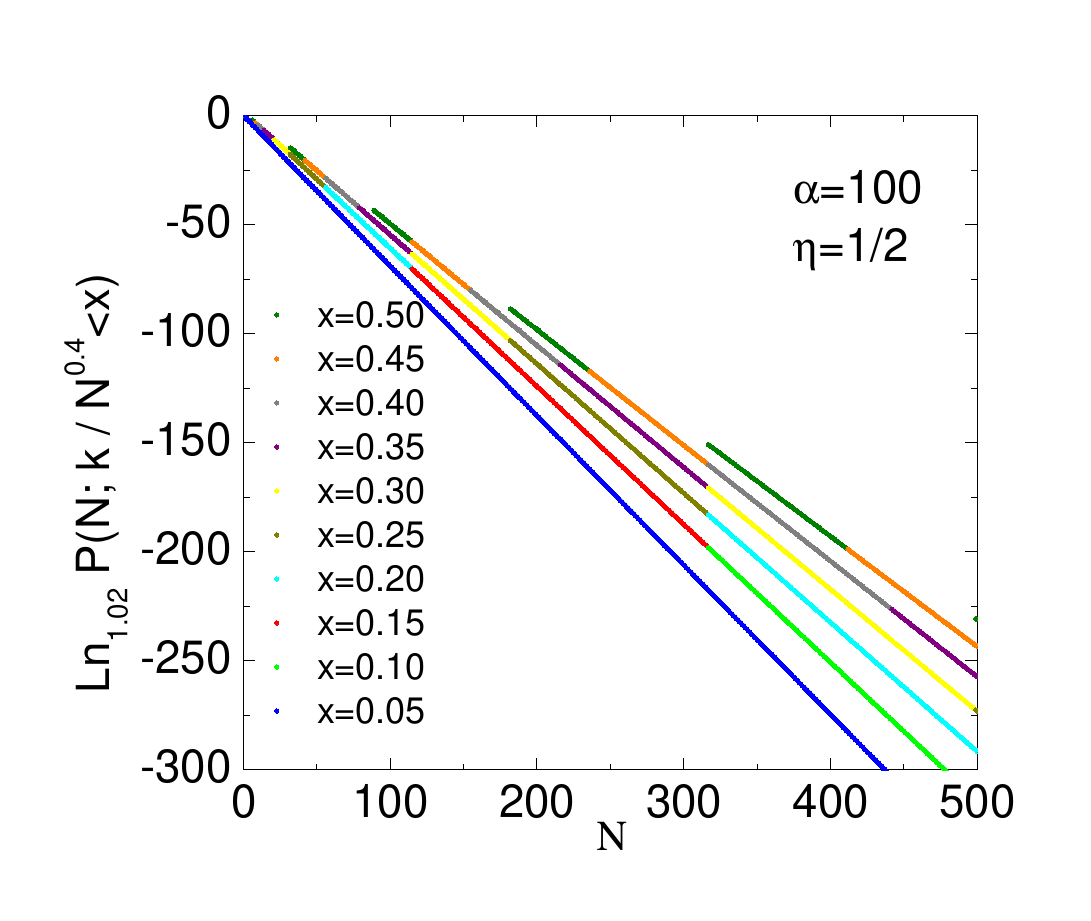}\hspace{-0.5cm}
\includegraphics[width=5.7cm]{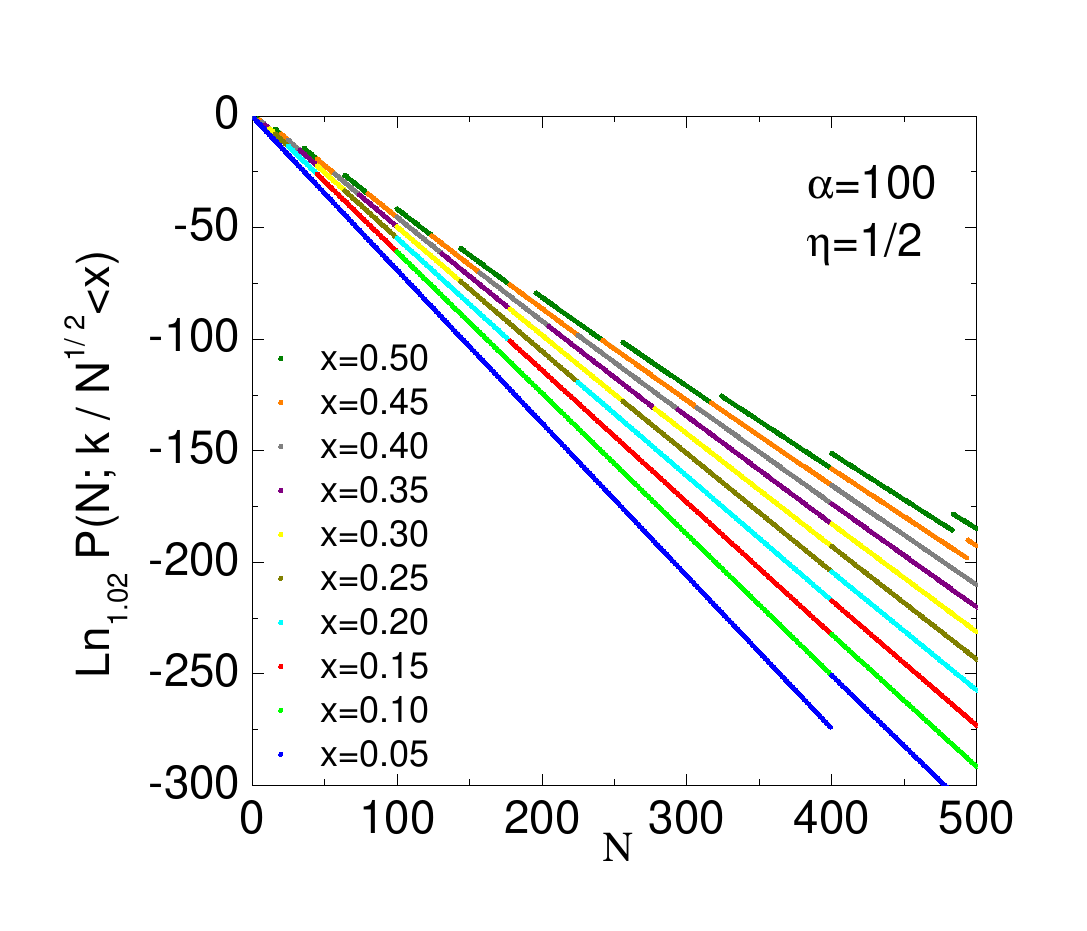}\hspace{-0.5cm}
\includegraphics[width=5.7cm]{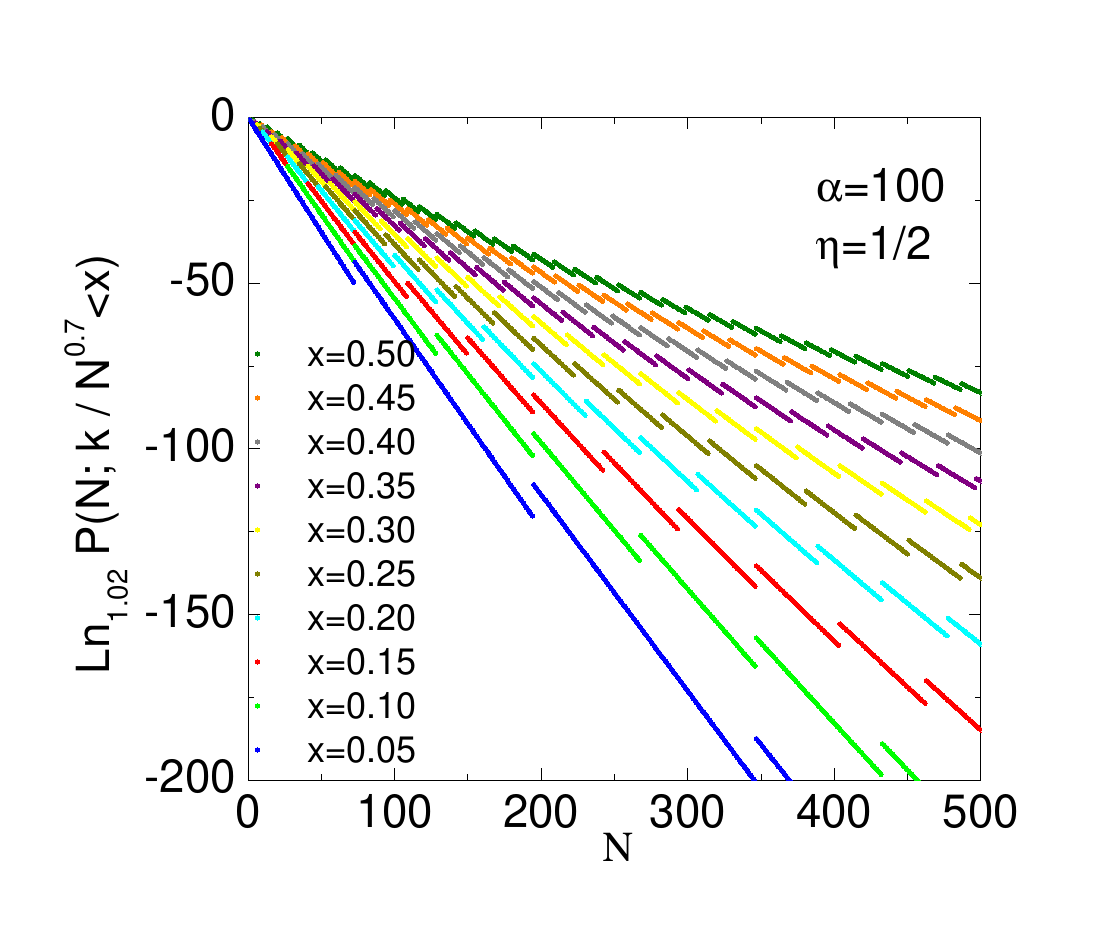}\\ \vspace{-0.5cm}
\includegraphics[width=5.7cm]{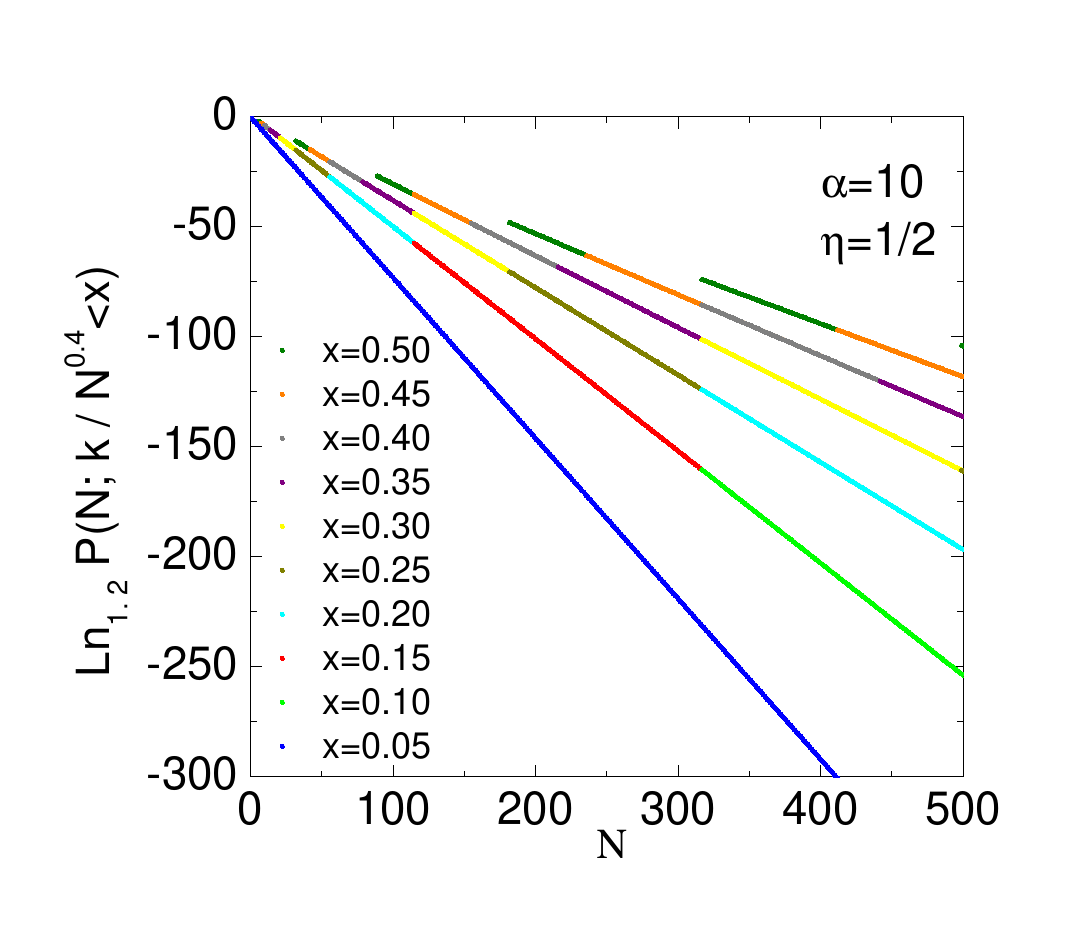}\hspace{-0.5cm}
\includegraphics[width=5.7cm]{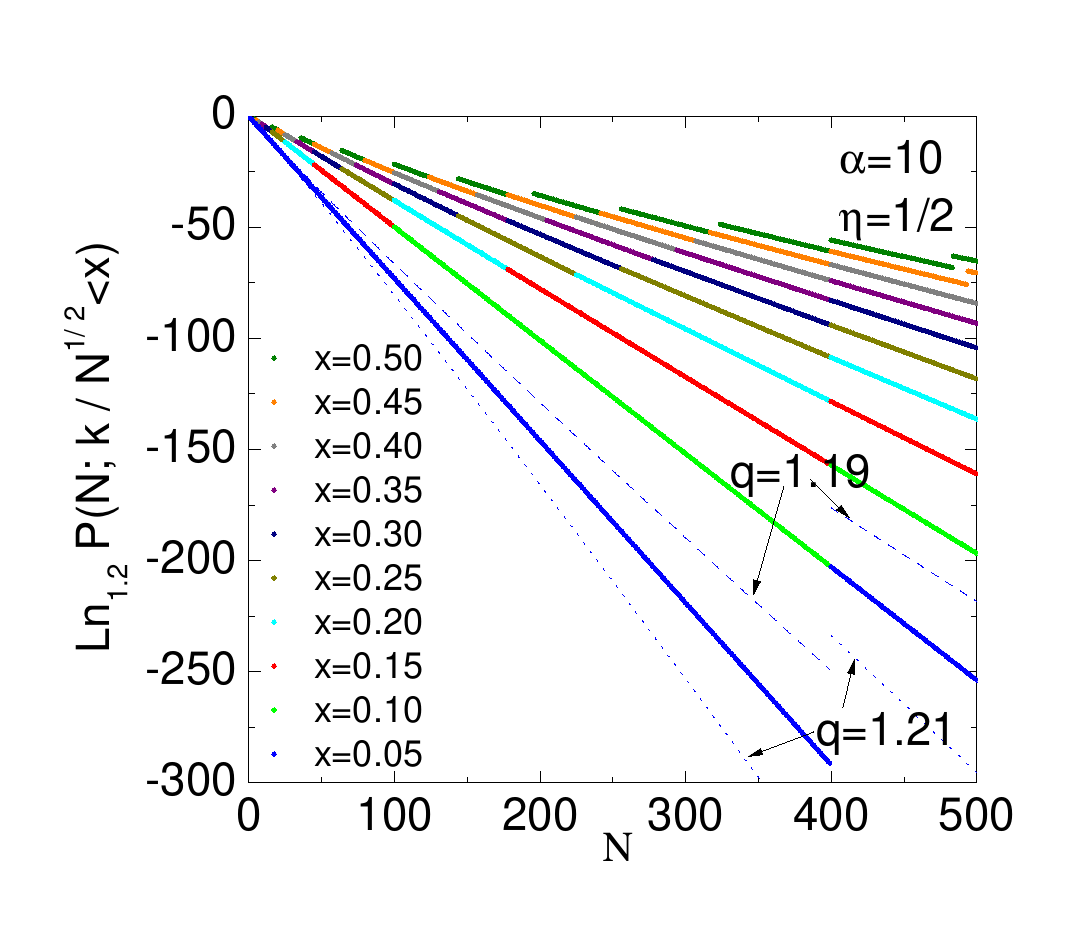}\hspace{-0.5cm}
\includegraphics[width=5.7cm]{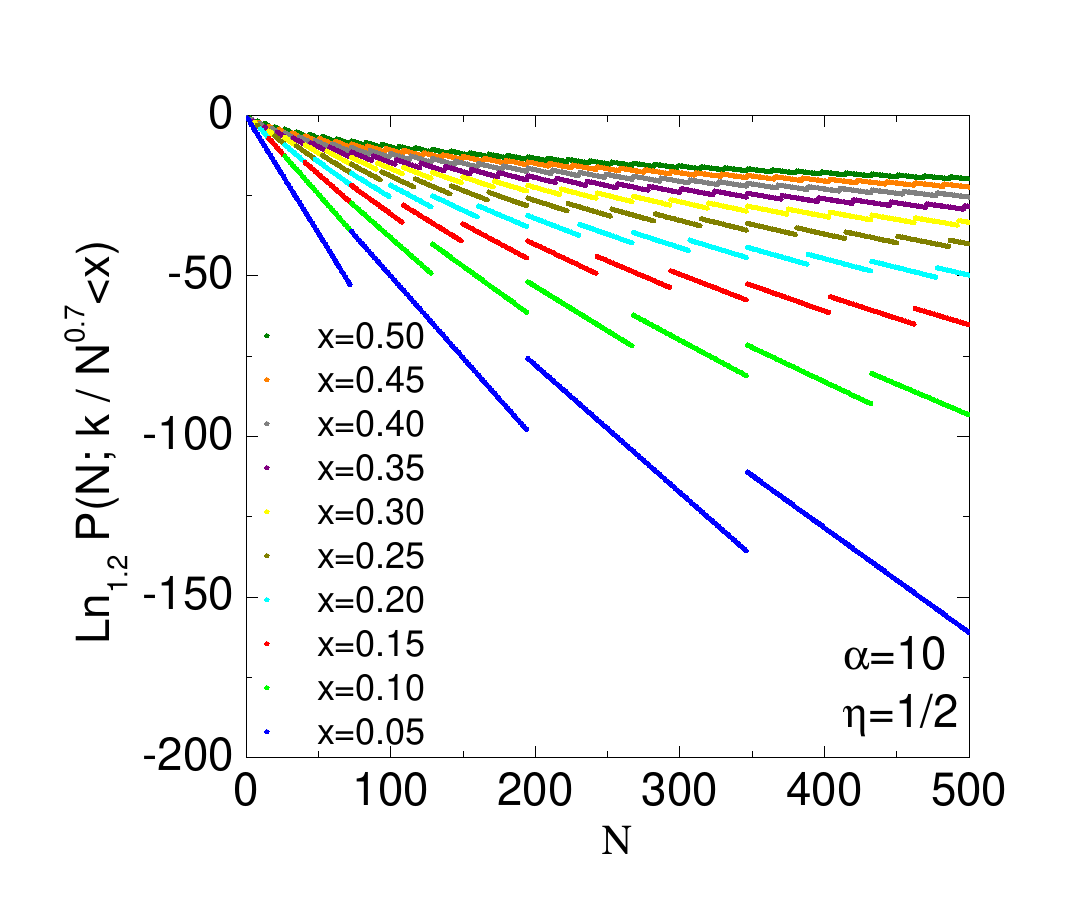}\\ \vspace{-0.5cm}
\includegraphics[width=5.7cm]{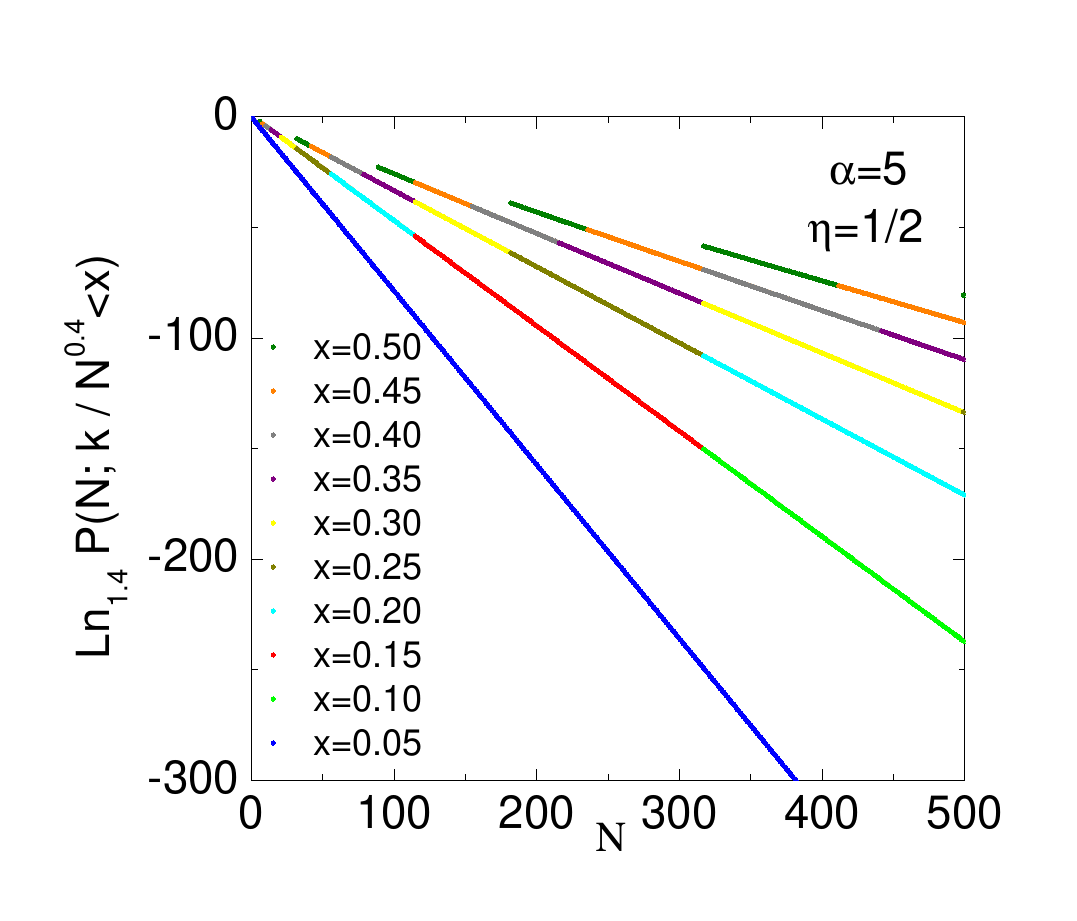}\hspace{-0.5cm}
\includegraphics[width=5.7cm]{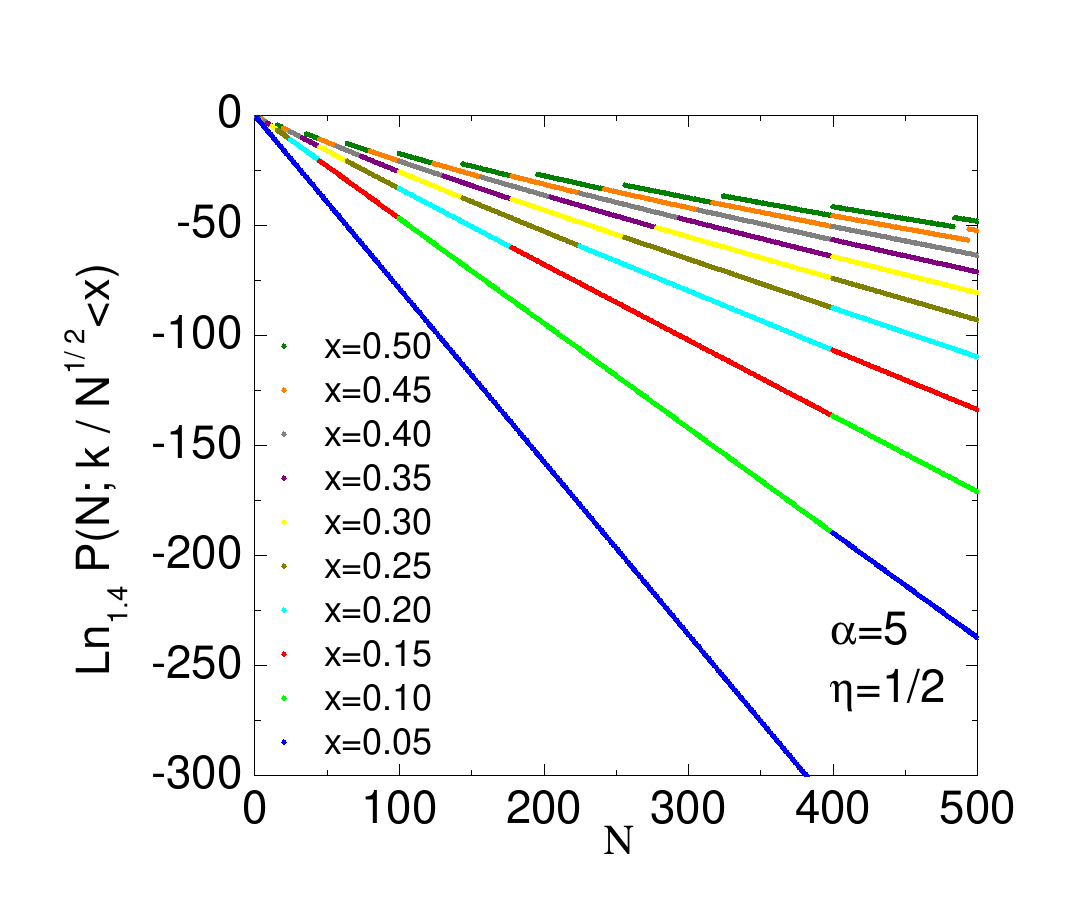}\hspace{-0.5cm}
\includegraphics[width=5.7cm]{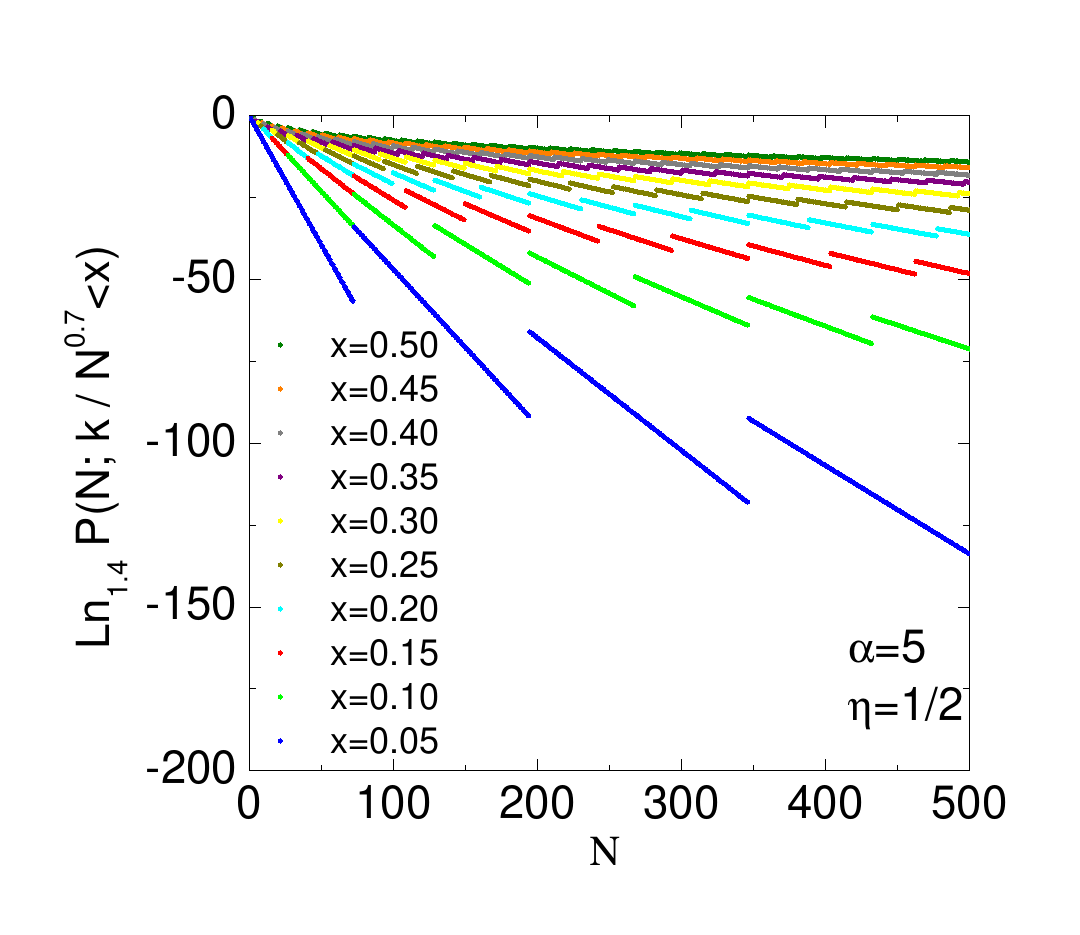}\\ 
\caption{$P(N; k/N^{\gamma}<x)$ distribution, for typical values of $\alpha$ ($\eta=1/2$), in semi-$q^{ldl}$-logarithmic representation. The values of $q^{ldl}(\alpha)$, related to the $q^{ldl}$-exponential decay, are  $q^{ldl}(100)=1.02$, $q^{ldl}(10)=1.2$ and $q^{ldl}(5)=1.4$, no matter the value of $\gamma$ ($\gamma=0.4$, $\gamma=0.5$ and $\gamma=0.7$, in figure). The central panel shows the significative deviations from a straight line, for  $1\%$ deviations of  $q^{ldl}$. \label{qLog}}
\end{center}
\vspace{-0.5cm}
\end{figure}

In the case  $\gamma<1$, we have numerically found a $q^{ldl}$-exponential  (where $ldl$ stands for {\it large-deviation-like}) decaying behavior  of the probability $P(N; k/N^\gamma\le x)$, for typical values of $\eta$ and $\alpha$, when $N$ increases. In fact (see Fig.~\ref{Histgama}) it can be straightforwardly verified that
\begin{equation}
\lim_{N\to \infty} P(N; \frac{k}{N^\gamma}\le x)=\lim_{N\to \infty} \sum_{k=0}^{\lfloor N^\gamma x \rfloor}\mathfrak{p}^{(N)}_k=0,
\end{equation}
and the zero limit is attained as
\begin{equation}
P(N; \frac{k}{N^\gamma}<x)\sim e_q^{-Nr_{q^{ldl}}}=\Bigl[1-(1-q^{ldl})Nr_{q^{ldl}}\Bigr]^{1/(1-q^{ldl})},
\end{equation}
with $q^{ldl}>1$. This result can be checked in Fig.~\ref{qLog}, where $q^{disc}(N)$-logarithmic representation of $ P(N; \frac{k}{N^\gamma}<x)$ as a function of $N$ is shown, for typical values of $\alpha$ and $\gamma$. Straight lines are obtained  for all values of $ \eta$ ($0\le \eta\le 1 $), $\alpha$ ($\alpha > 0$) and $\gamma$ ($\gamma <1$) that have been considered.

\begin{figure}
\begin{center}%
\includegraphics[width=7cm]{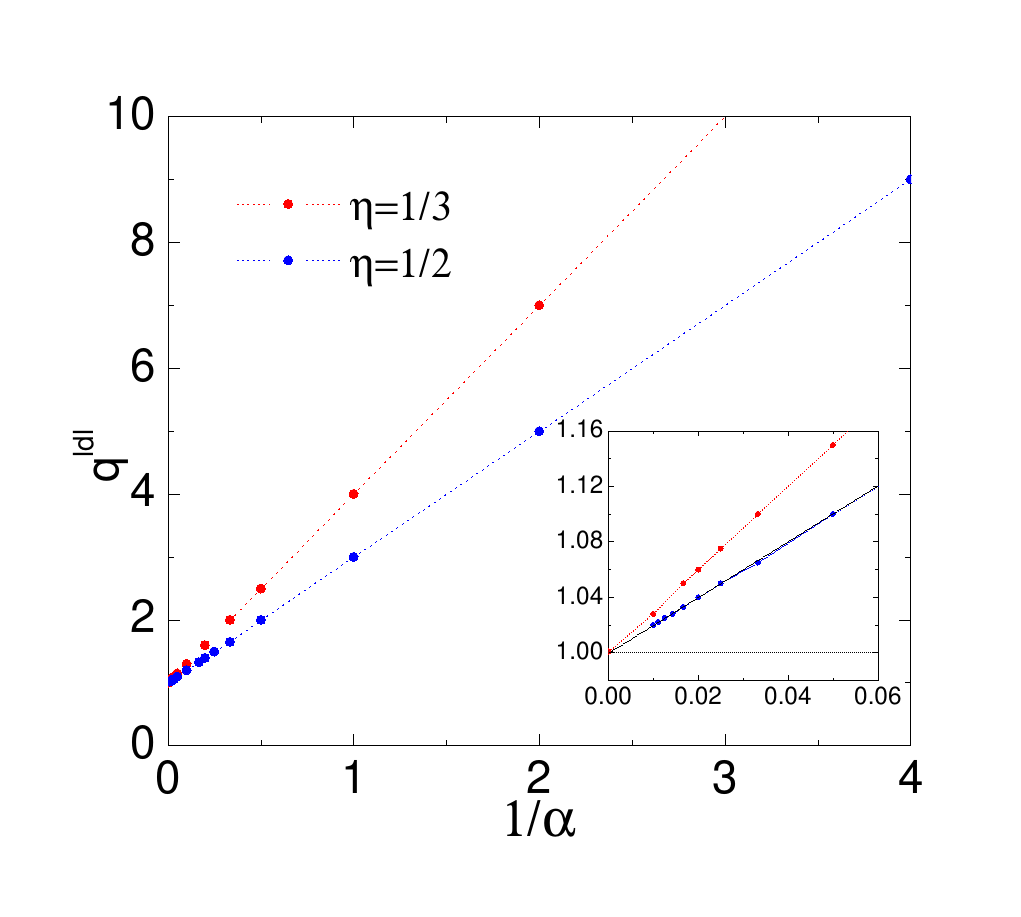}
\end{center}
\caption{The $\alpha$-dependence of the $q^{ldl}$-exponential  index of the decaying  probability $P(N; k/N^{1/2}<x)$, for   $\eta=1/2$ and $\eta=1/3$  models.
\label{qalpha}}
\end{figure}
\begin{figure}
\begin{center}\vspace{-0.5cm}
\includegraphics[width=7.5cm]{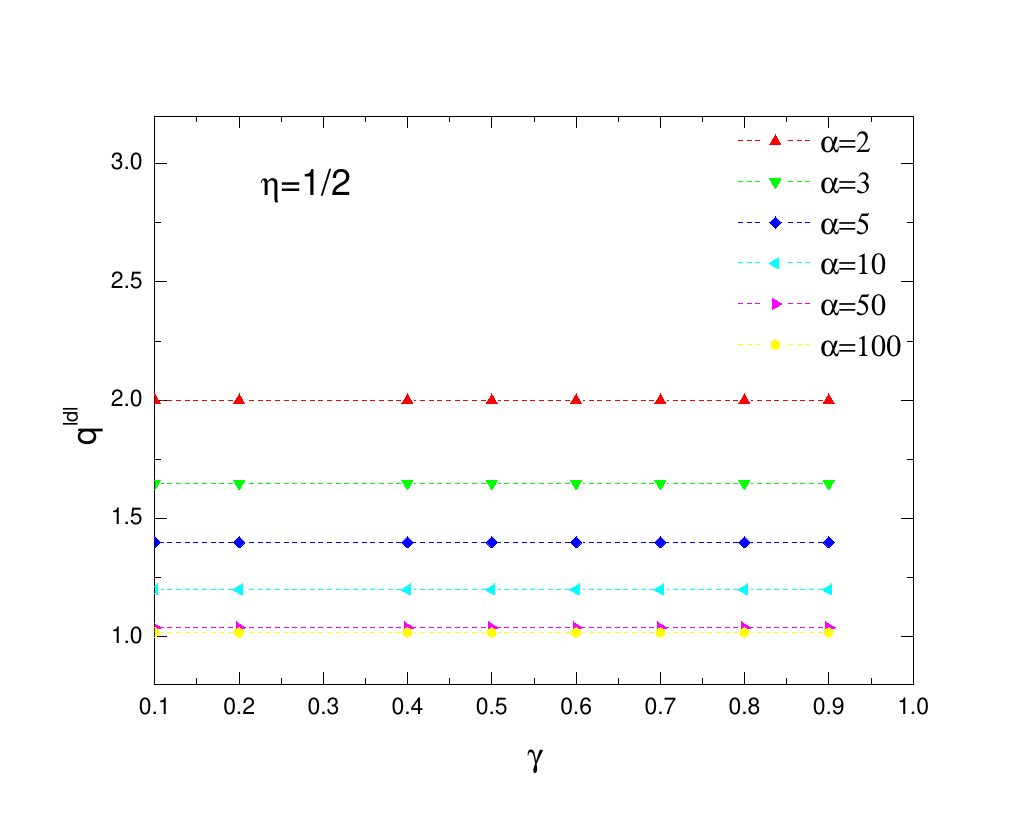}\vspace{-0.5cm}
\end{center}
\caption{The $q^{ldl}$-exponential  indices appear to be independent of $\gamma$. \label{qgamma}}
\end{figure}
We have  heuristically obtained (see Fig.~\ref{qalpha}) that, for all the values of $x$ for which we have checked, the $q^{ldl}$ index  satisfies
 \begin{equation}
q^{ldl}(\eta, \alpha)=1+\frac{1}{\eta \alpha},\label{qldtetaalfa} \,.
 \end{equation}
Notice that it does not depend on $\gamma$ or $x$ (see Fig.~\ref{qgamma}).

Consequently, from (\ref{qalfa}) and (\ref{qldtetaalfa}), we  infer that, for $\eta=1/2$ and for all values of $\alpha>0$,   the $q^{ldl}(\alpha)$-exponential decay index and  the $q^{att}(\alpha)$-Gaussian attractor index are related as follows:
 \begin{equation}
 \left.\frac{1}{q^{att}( \alpha)-1}+\frac{1}{q^{ldl}(\alpha)-1}\right|_{\eta=1/2}=1.
  \end{equation}

\begin{figure}
\begin{center}
\includegraphics[width=7.5cm]{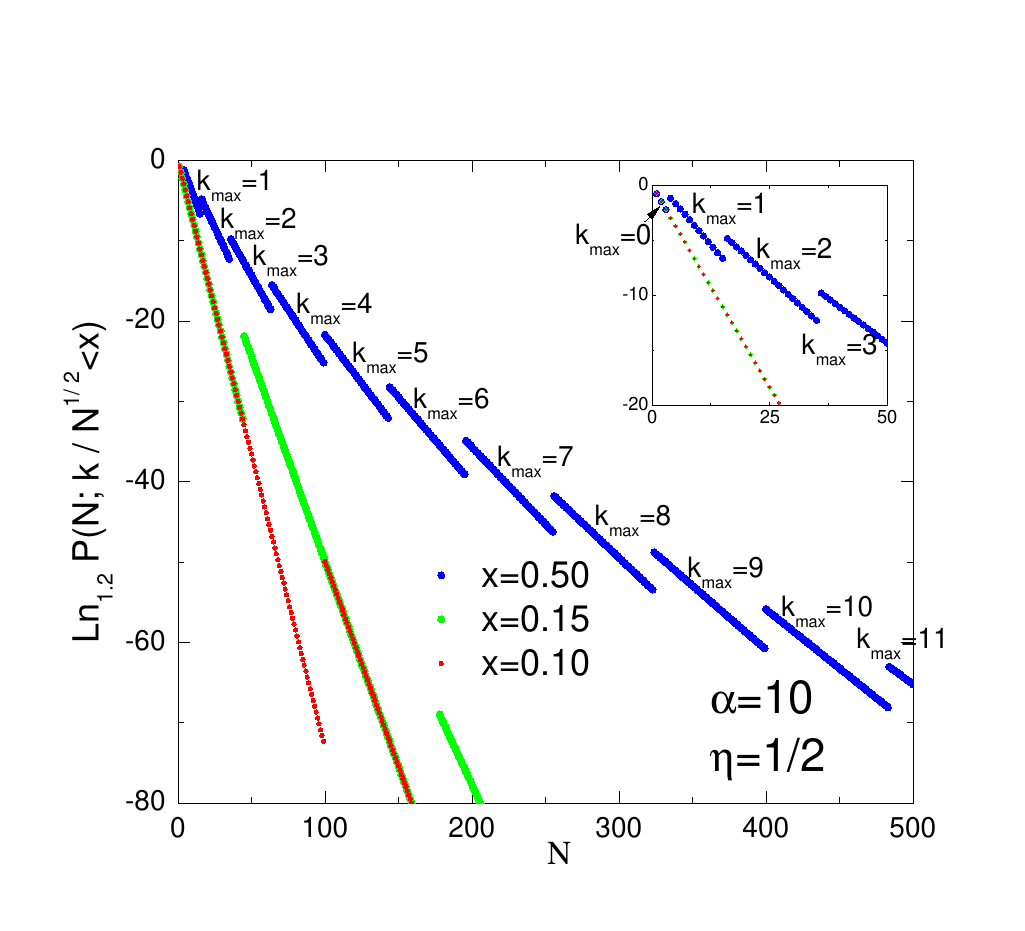}\vspace{-0.5cm}
\end{center}
\caption{$q^{ldl}$-exponential decaying rate of $P(N;k/N^{1/2}<x)$ for $\eta=1/2$ and $ \alpha=10$  shows that the sequence of slopes  that correspond to a value of $x$, are associated to  $k_{max}\equiv \max\{\lfloor N^{\gamma} x \rfloor$\} involved   in  $P(N; k/N^\gamma <x)$.\label{qexpx}}
\end{figure}

Let us now focus on the values of the slopes of the semi-$q^{ldl}$-logarithmic representation of $P(N; k/N\le x)$.  Fig.~\ref{qLog} shows that the  $q^{ldl}$-exponential decaying rate does not only depend on the value of $x$, i.e., $r_{q^{ldl}}=r_{q^{ldl}}(x, N, \gamma;\eta, \alpha)$.
The mechanism that precludes
a simple dependence on $x$ can be understood by fixing a particular value of $x$ and $\gamma$, as shown in Fig.~\ref{qexpx}.
For a fixed value of $x$, the sequence of slope values are associated to the maximum value of $k$ involved in $P(N;k/N^{\gamma}<x)$, i.e., $k_{max}(x,N,\gamma)=
\lfloor N^{\gamma} x \rfloor$. Summarizing, the deviation re-scaled probability behavior presents a $q^{ldl}$-exponential decay  when $\gamma < 1$, that can be written as
\begin{equation}
P(N; k/N^\gamma <x;\eta, \alpha)=\sum_{k=0}^{
\lfloor N^{\gamma} x \rfloor}\mathfrak{p}^{(N)}_k(\eta, \alpha)\simeq e_{q^{ldl}=1+\frac{1}{\eta \alpha}}^{-Nr(\lfloor N^{\gamma} x \rfloor; \, \eta, \alpha)} \,,
\end{equation}
thus exhibiting a non-trivial dependence of the rate function $r_{q^{ldl}}(\lfloor N^{\gamma} x \rfloor ; \, \eta, \alpha)$.
\section{Large deviation probability with respect to the $N\to \infty$ limit distribution}
\label{sec5}
Let us now analyze the $N\to \infty$ evolution behavior of the probability of  $k/N^\gamma$ for $\gamma=1$, i.e., how $p(k/N)$ approaches its  attractor ${p}_{\infty}(x)$.
The probability left deviations of $k/N$ from $x$ ($0\le x\le 1/2$), $P(N;k/N<x)$,  with respect to the  attractor, can be written as
\begin{equation}
\mathbb{P}_k^{(N)}(x; \eta, \alpha)\equiv P\left(k/N\le x;  \eta, \alpha\right)-
 P^{att}\left(x;  \eta, \alpha\right),
\end{equation}
where $P^{att}\left(x;  \eta, \alpha\right)\equiv\lim_{N\to \infty}
P\left(k/N\le x;  \eta, \alpha\right)=
\int_0^xp_{\infty}(z; \eta, \alpha)dz$.

From Eq.~\eqref{improvenum}, we obtain that the  probability left deviations of $k/N$ from $x$, for  $\eta=1/2$ and even values of $\alpha\ge 4$ (i.e. $4\le \alpha=\dot{2}$), can be writen as
\begin{equation}
P\left(N; k/N\le x; \alpha\right)=\sum_{k=0}^{\lfloor N x \rfloor}\mathfrak{p}^{(N)}_k(1/2, \alpha)=\mathfrak{p}^{(N)}_0(1/2, \alpha)+\frac{N!}{(\alpha)_N(\alpha/2-1)!^2}\sum_{k=1}^{\lfloor N x \rfloor}\prod_{j=1}^{\alpha/2-1}
(k+j)
(N-k+j) \,.
\end{equation}

Let as analytically study the simplest model, i.e., $\eta=1/2$ and $\alpha=4$.
We verify
\begin{eqnarray}P\left(N;k/N\le x\right)&=& \frac{6}{(N+2)(N+3)}+\frac{N!}{(4)_N}\sum_{k=1}^{\lfloor N x \rfloor}
(k+1)(N-k+1)\nonumber\\
 &=&\frac{6(N+1)+\lfloor N x \rfloor(5+9N+3\lfloor N x \rfloor(N-1)-2(\lfloor N x \rfloor)^2}{(3+N)(2+N)(1+N)}.\label{DistbdevN}\end{eqnarray}
The corresponding  $N\to \infty$ limit distribution is a $q^{att}$-Gaussian, with  $q^{att}=0$. Consequently, the corresponding asymptotic left  deviation of $k/N$ from a fixed value $x$ is given by
\begin{equation}
P^{att}\left(x\right)= \int_0^xp_{\infty}(z)dz
=\frac{2 \, \Gamma\left(\frac{5}{2}\right)}{\sqrt{\pi} \, \Gamma(2)}\int_0^x
\left[1-4 \left(X-\frac{1}{2}\right)^2\right]dX=
3x^2-2x^3.\label{AttdevN}
\end{equation}
From \eqref{DistbdevN} and  \eqref{AttdevN} we have that, for a certain value of $N$,  the  probability left deviations of $k/N$ from $x$, with respect to the  probability left deviation  of the $N\to \infty$  limit distribution, is analytically obtained as
{\small \begin{equation}
\mathbb{P}_k^{(N)}(x)=\frac{6(N+1)+\lfloor N x \rfloor(5+9N+3\lfloor N x \rfloor(N-1)-2(\lfloor N x \rfloor)^2}{(3+N)(2+N)(1+N)}-3x^2+2x^3.
\label{alfa4accdisttib}
\end{equation}}
The upper bound
$\Delta^{upp}_N(x)$ of \eqref{alfa4accdisttib} can be obtained in the case that  $ \lfloor N x \rfloor = N x$, as shown in Fig.~\ref{ldpatt}.
A lower bound $\Delta^{low}_N(x)$ can be also considered as $ N x-1< \lfloor N x \rfloor $, but it is never attained and $\Delta^{low}_N(x)< \mathbb{P}_k^{(N)}(x; \eta, \alpha)$. We can consider, instead of $\Delta^{low}_N(x)$, the maximum of all lower bounds  $\Delta^{min}_N(x)$, for each value of $x$. All these bounds verify the  relation
\begin{equation}
\Delta^{low}_N(x)<\Delta^{min}_N(x)\le \mathbb{P}_k^{(N)}(x; \eta=\frac{1}{2}, \alpha=4) \le \Delta^{upp}_N(x)
\end{equation}

\begin{figure}
\begin{center}
\includegraphics[width=7.1cm]{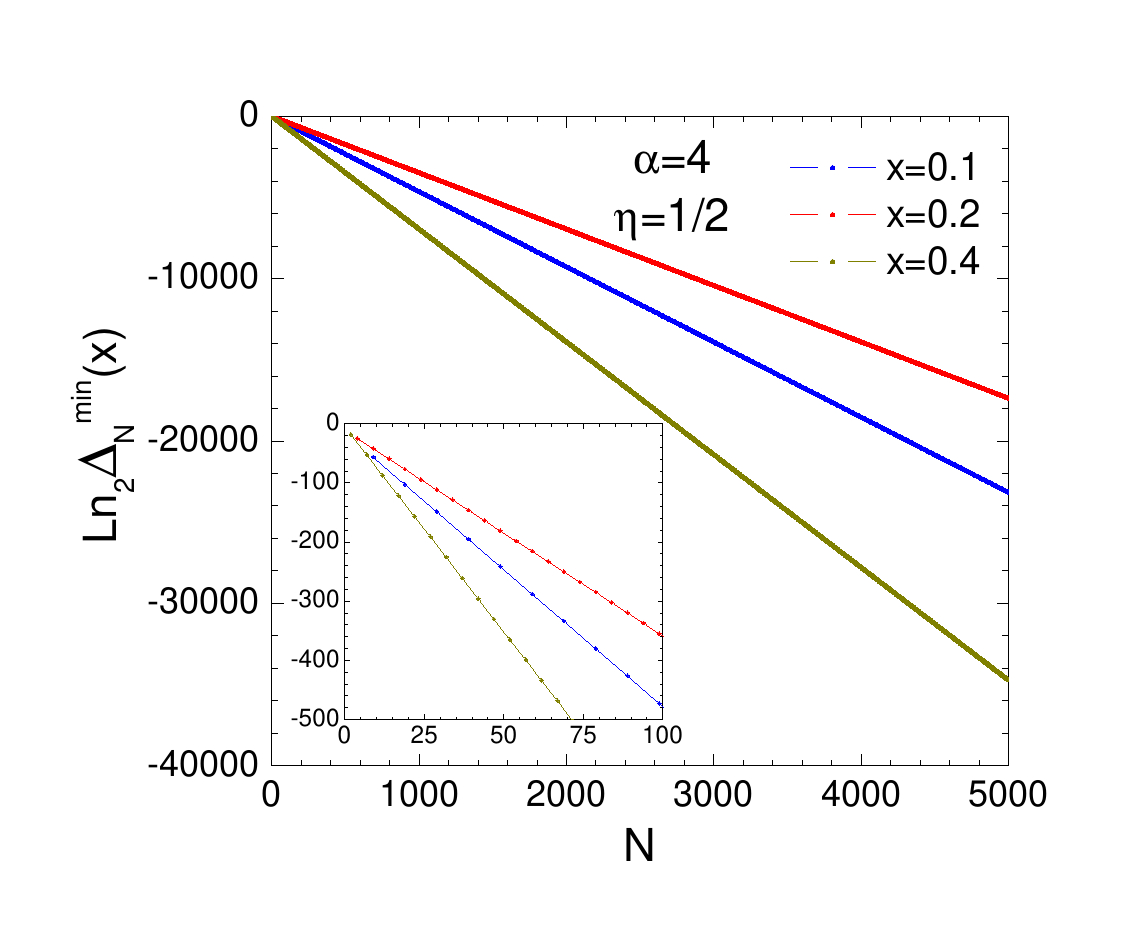}
\includegraphics[width=6.9cm]{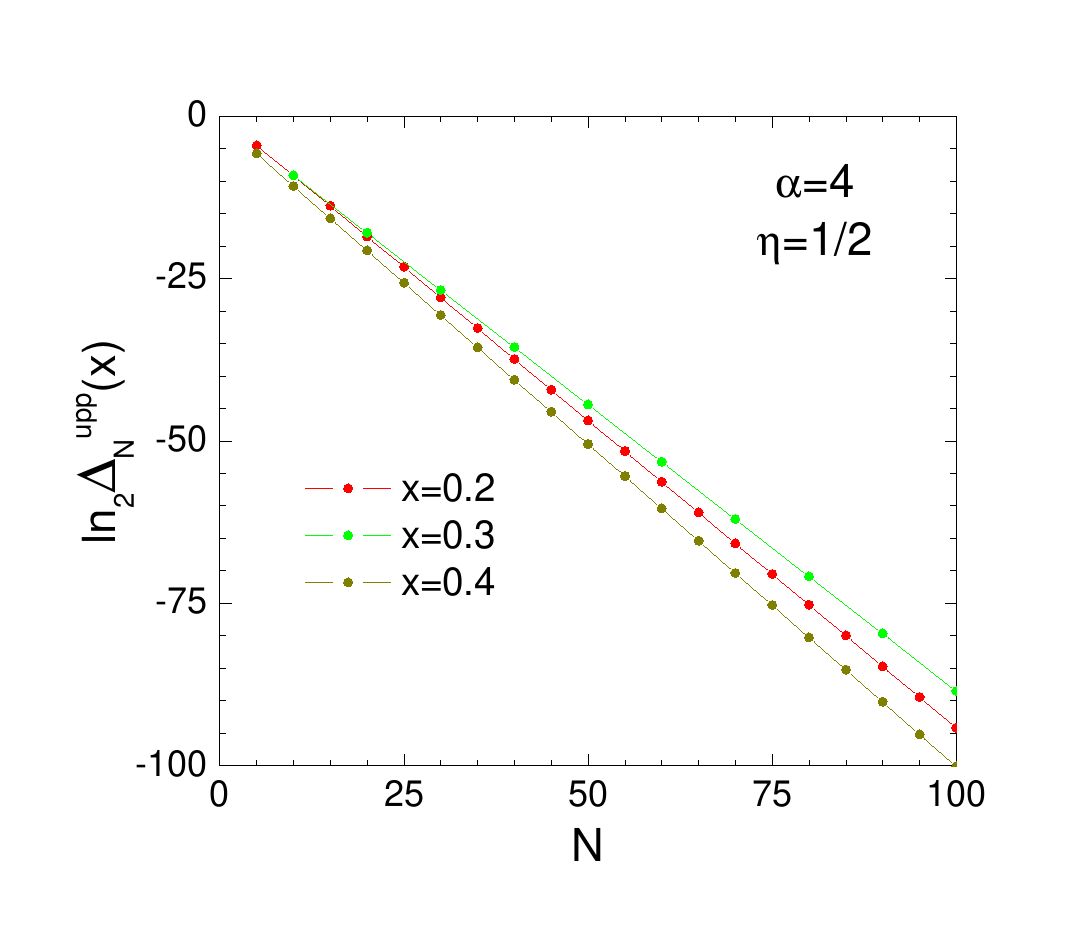}
\end{center}
\caption{{\it Left Panel:} Semi-$(q^{LD}=2)$-logarithmic representation  of $\Delta^{min}_N(x)$. Observe that, for $x=0.1$, a bias from the linear behavior exists.
{\it Right Panel:} The linear behavior of the $(q=2)$-logarithmic representation of the upper bound $\Delta^{upp}_N(x)$  could reflect its $q^{LD}$-exponential decay with $N$.  \label{qexpUpp}}
\end{figure}
Fig.~\ref{qexpUpp} exhibits the $(q=2)$-logarithmic representation of the upper and the minimum bound deviations, $\Delta^{upp}_N(x)$ and $\Delta^{min}_N(x)$. $\Delta^{upp}_N(x)$ appears to $q^{LD}$-exponentially decay with $N$ (were $LD$ stands for {\it Large Deviation}), as conjectured in \cite{Ruiz13}. It is not the case of $\Delta^{min}_N(x)$ for $x=0.3$, as it is shown in Fig.~\ref{ldpatt}.

\begin{figure}
\begin{center}
\includegraphics[width=5.9cm]{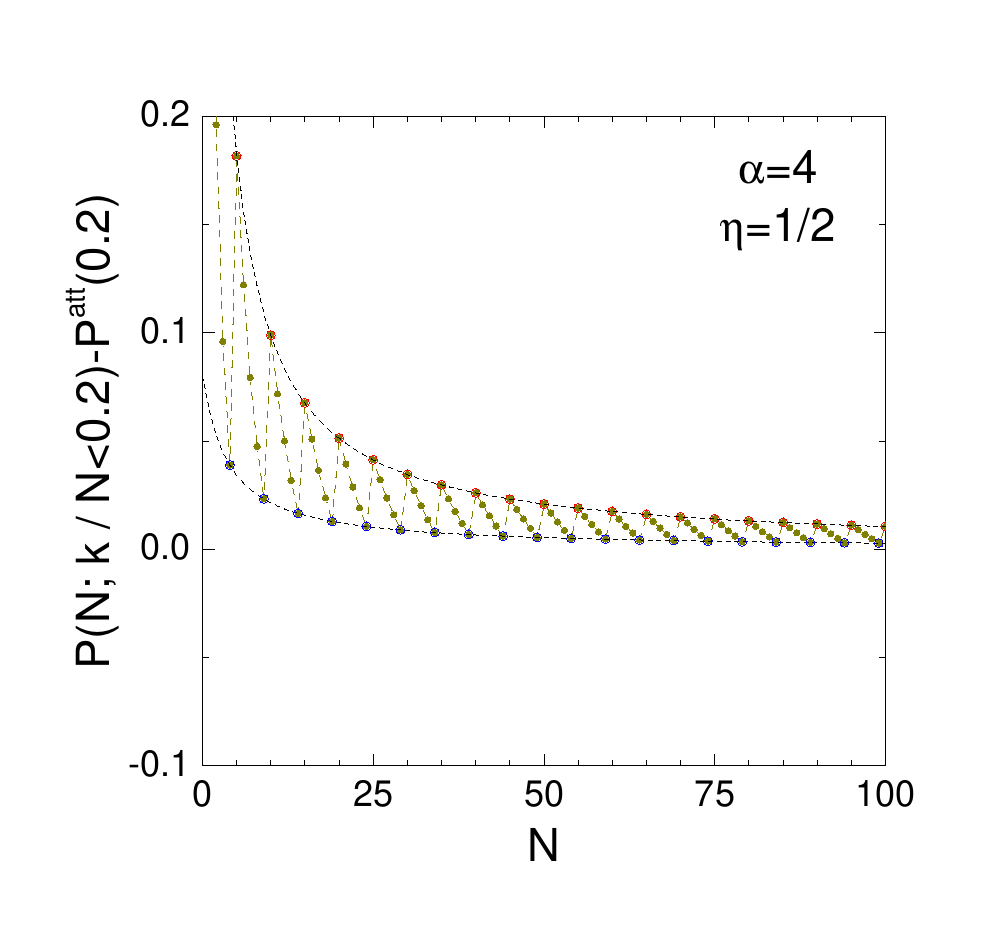}
\includegraphics[width=5.9cm]{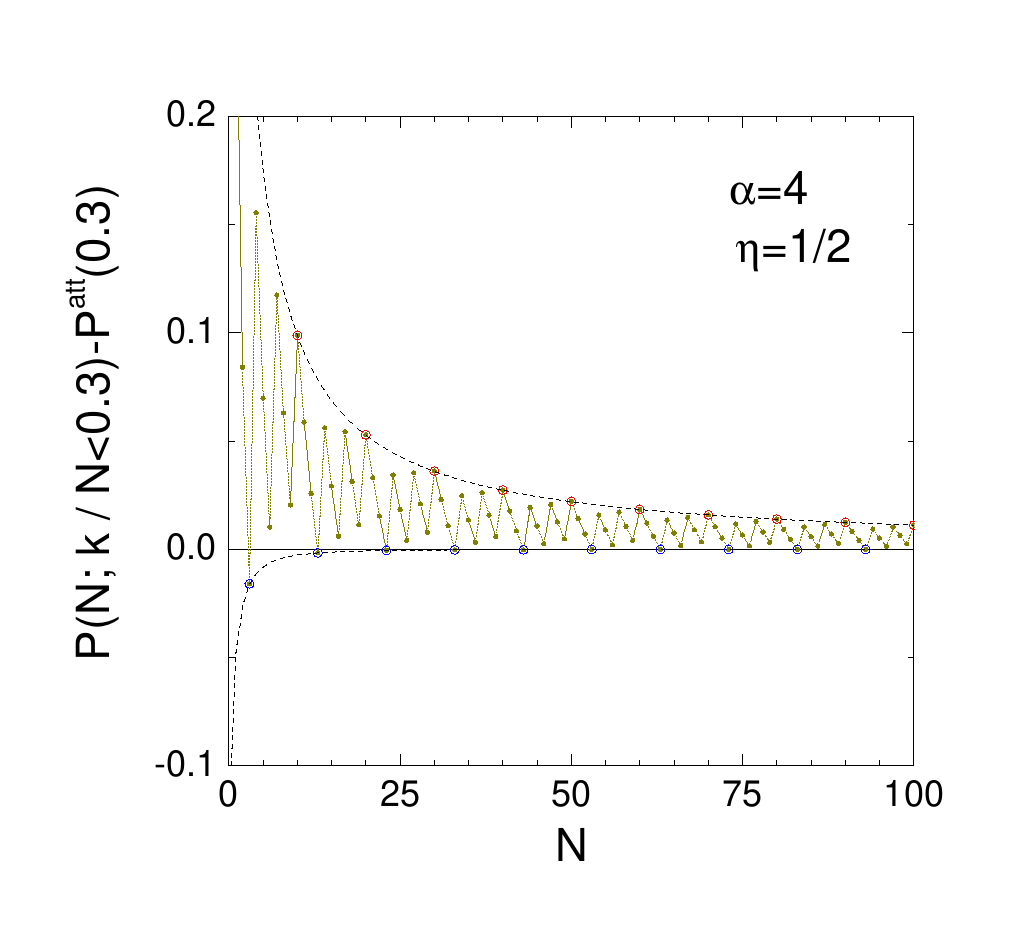}
\includegraphics[width=5.9cm]{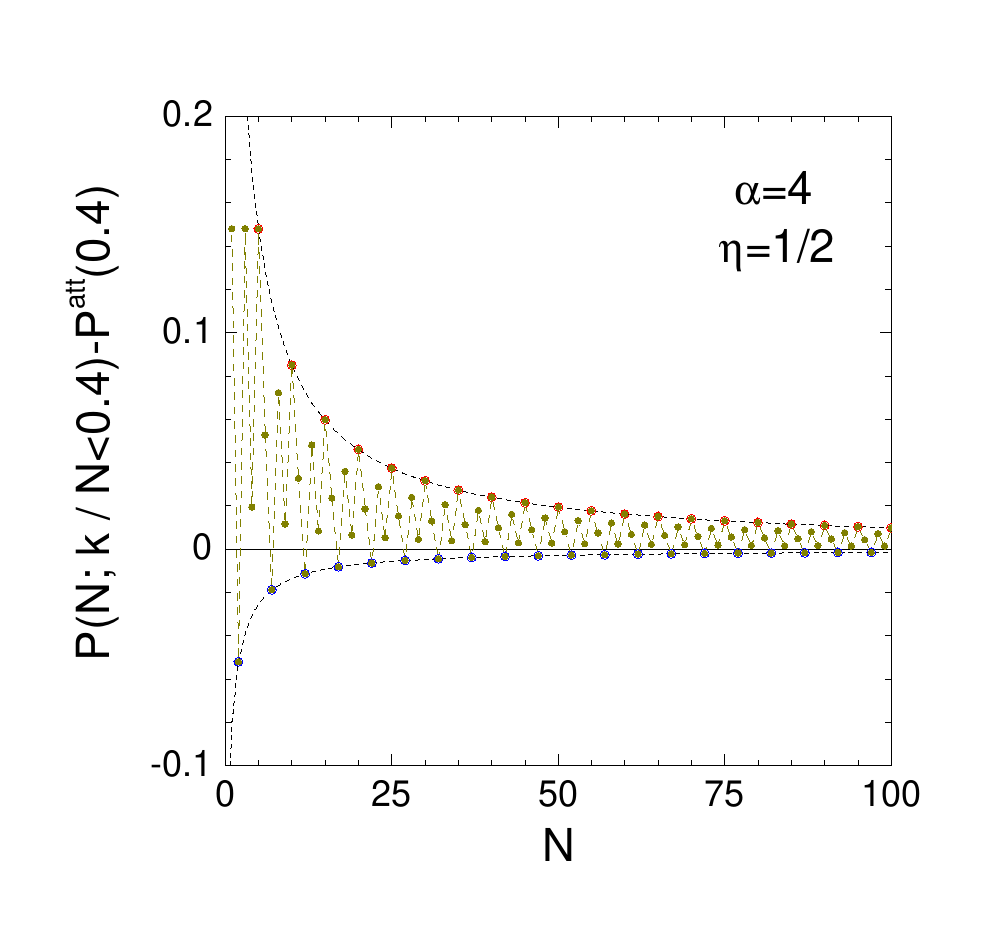}\\ 
\includegraphics[width=5.9cm]{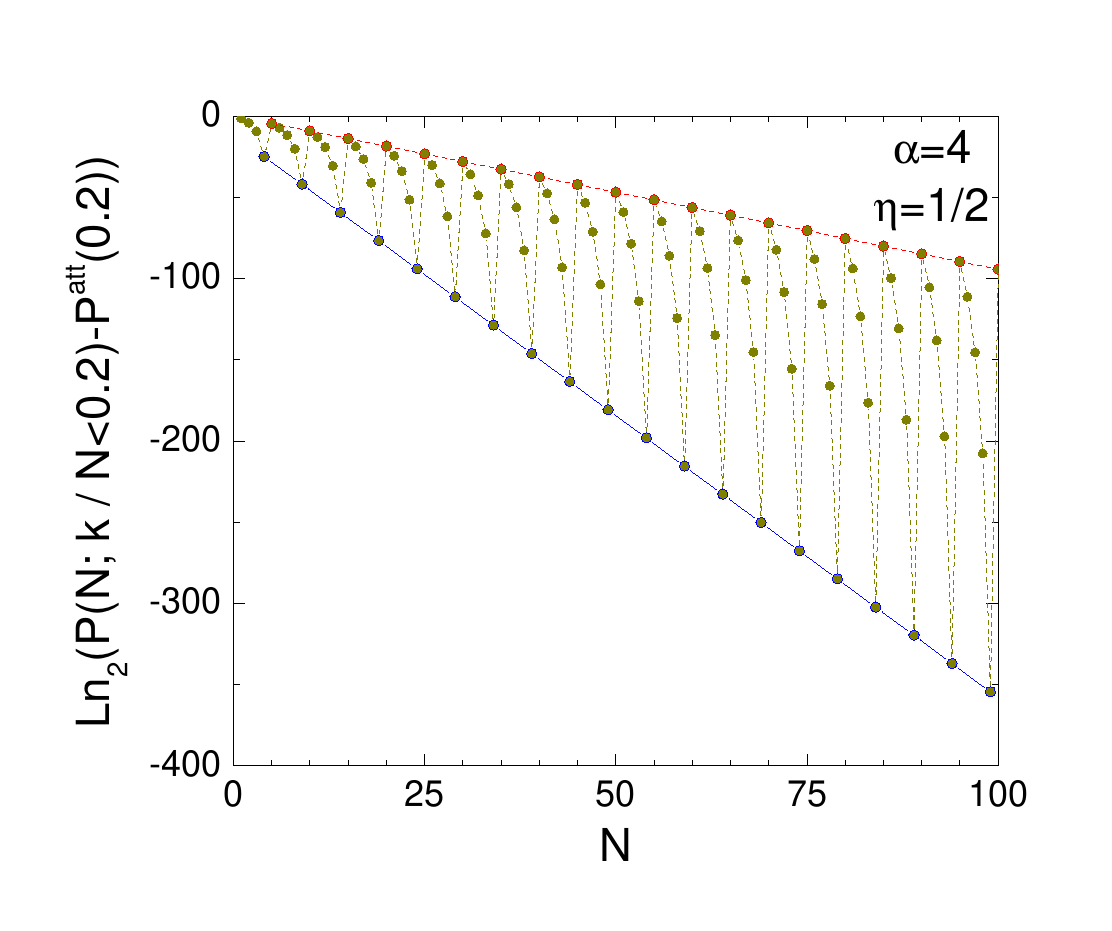}
\includegraphics[width=5.9cm]{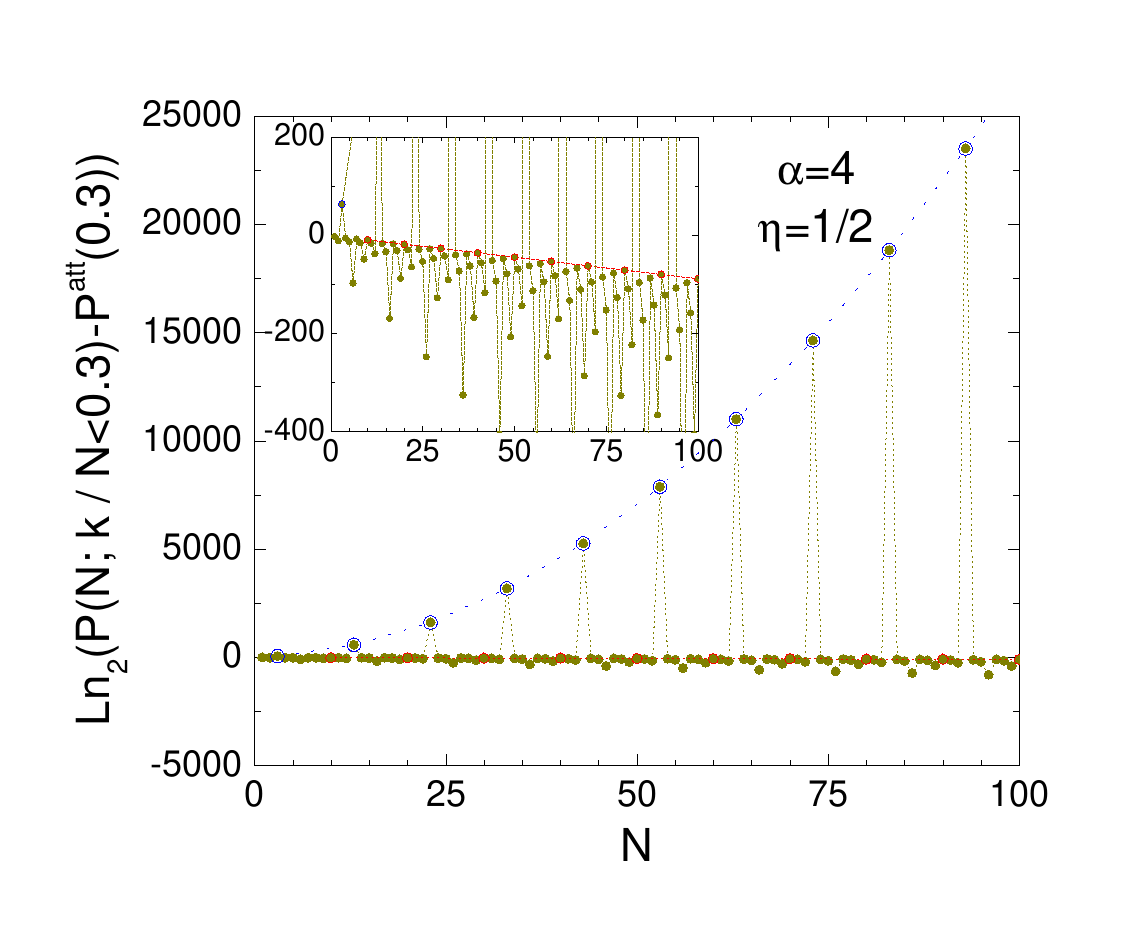}
\includegraphics[width=5.9cm]{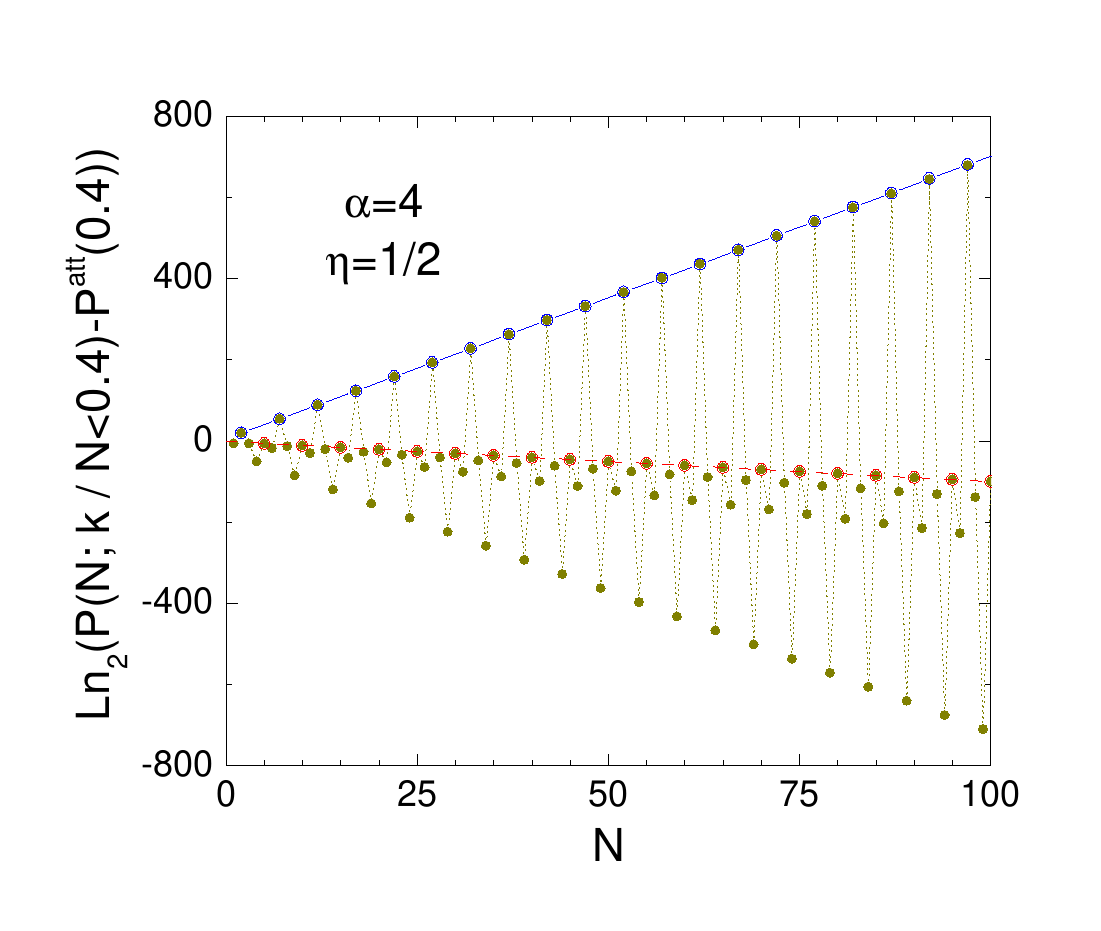}
\end{center}\vspace{-0.5cm}
\caption{Upper figures represent the probability of a left  deviation of $k/N$ from a fixed value of $x$, with respect to the deviation in $N\to \infty$ limit distribution, i.e., $\mathbb{P}_k^{(N)}(x; \eta, \alpha)=P(k/N\le x; \eta, \alpha)-P_\infty(z\le x; \eta, \alpha)$ for $\eta=1/2$ and $\alpha=4$. From left to right, $x=0.2$, $x=0.3$ and  $x=0.4$.
The upper and minimum bounding sets, i.e., $\Delta^{upp}_N(x)$ and $\Delta^{min}_N(x)$, are dashed in red and blue, respectively. Bottom figures show the same results in semi-$q^{LD}$-logarithmic representation, where $q^{LD}=2$ in all cases.\label{ldpatt}}
\end{figure}
The  hypothesis of $\Delta^{upp}_N(x)$ $q^{LD}$-exponentially decaying behavior can  be verified by using the asymptotic expansion of the analytical expressions of  the bounding values $\Delta^{upp}_N(x)$, $\forall x$:
\begin{equation}
\Delta^{upp}_N(x)= \frac{3x(x-1)(4x-3)}{N}\left[1-\frac{50x^2-43x+6}{3x(4x-3)N}
+\frac{5(3x-2)(4x-1)}{x(4x-3)N^2}+\dots\right] \,.
\label{aimpt}
\end{equation}
We can compare the respective terms with the asymptotic expansion of a $q$-exponential
function \cite{Ruiz13}, namely
{\small
\begin{eqnarray}
a(x) e^{-r_q(x)N}_q&=&\frac{a(x)}{[(q-1)r_q(x)N]^{\frac{1}{q-1}}}\nonumber\\
& \times & \left[
1-\frac{1}{(q-1)^2r_q(x)N}+\sum_{m=2}^{\infty}(-1)^m\frac{q(2q-1)\dots[(m-1)q-(m-2)]}{m!(q-1)^{2m}(r_q(x)N)^m}
\right]\,. \label{qexpesp}
\end{eqnarray}}

Equations (\ref{aimpt}) and (\ref{qexpesp}) would provide, by neglecting higher-order terms, an index $q^{LD}=2$. In such a situation, and identifying the two first terms of expansions, the best  $q^{LD}$-generalized rate function and the corresponding  $q^{LD}$-exponential factor $a(x)$ would be
\begin{equation}
r_q^{upp}=\frac{3x(4x-3)}{50x^2-43x+6} \,, \qquad
a^{upp}(x)=\frac{9x^2(x-1)(4x-3)^2}{50x^2-43x+6}.\label{qupp}
\end{equation}
But, in fact, the third term of Eq.~\eqref{qupp} is not negligible for some  values of $x$ and, in such cases,  Eq.~\eqref{qexpesp} and  Eq.~\eqref{qupp} are not compatible.
The $q^{LD}$-exponential decay of the upper bound of the large deviation to the attractor is precluded, as  obtained within a different context in \cite{Max}.

Other values of $\alpha$ have been tested and, in all cases, the large deviation probability with respect the attractor presents  a power-law decay. The large-deviation probability does not in fact $q$-exponentially decay to the attractor,  even though, in some cases,     Eq.~\eqref{qexpesp} roughly  describes the large-deviation behavior.

\section{Conclusions}
\label{Conclusions}
A generalized binomial distribution based on $q$-exponential generating functions is characterized by Eqs. (\ref{distribevaldo}) and (\ref{improvenum}). Its  probability function $\mathfrak{p}^{(N)}_k(\eta, \alpha)$ depends on two parameters $(\eta, \alpha)$, and can be considered as the following urn scheme: from a set of $b$ black balls and $r$ red balls contained in an urn, one extracts one ball and returns it to the urn, together with $c$ balls of the same color. In that case,
$\mathfrak{p}^{(N)}_k(\eta, \alpha)$ represents the  probability to have  $k$ black balls in the urn after the {\it N}-th trial, and it can   can be written as a function of $(b,r,c)$, as $\eta=b/(b+r)$ and $\alpha=(b+r)/c$ \cite{Polya}.

If  no bias exists, i.e. for $\eta=1/2$, the probability to find a relative number of black balls $k/N$ after  the {\it N}-th trial, closely approaches  a $q^{disc}(N)$-Gaussian distribution. The $N\to \infty$ limit probability distribution is in fact a $q^{att}$-Gaussian \cite{Evaldo2014} whose $q^{att}$ index and $\beta^{-1}$  generalized temperature, can be obtained  from $(b,r,c)$ as
$q^{att}=\frac{b+r-4c}{b+r-2c}$ and $\beta^{-1}=\frac{c}{2(b+r-2c)}$.
In other words,
the numerical discrete distributions  appear to be very close to a set of $q^{disc}(N)$-Gaussian distributions that, increasing $N$, evolve towards to a $q^{att}$-Gaussian attractor. This urn scheme provides a procedure to attain a $q^{att}$-Gaussian $N\to \infty$ limit distribution that verifies, in all cases ($\forall a,b,c$),  the relation  $\beta(1-q^{att})=4$. When the number of reposition balls verifies $c<(b+r)/2$, such attractors are  concave $q^{att}$-Gaussian distributions with a bounded and compact support (where index $q^{att}<1 $ and generalized temperature $\beta^{-1}>0$); on the contrary, when $c>(b+r)/2$, such  attractors are convex $q^{att}$-Gaussian distributions with  bounded but non compact support (where index $q^{att}>2 $ and  generalized temperature $\beta^{-1}<0$).

These generalized binomial distributions violate the law of large numbers, but nevertheless present a large-deviation-like property. Indeed, by using, instead of the variable $k/N$, the rescaled variable $k/N^{\gamma}$ ($\gamma<1$), the  left deviation  probability behaves as  $P(N; k/N^\gamma <x;\eta, \alpha)\simeq e_{q^{ldl}}^{-Nr(\lfloor N^{\gamma} x \rfloor; \, \eta, \alpha)}$. Moreover, an interesting result is that, when no bias exists (i.e., $\eta = 1/2$), the $q^{att}$-Gaussian index, the $q^{ldl}$ index, and the $q^{gen}$ index (characterizing the generating function), are univocally defined by the $(b+r)/c$ ratio, and simple mathematical relations exist that remind the  algebra indicated in \cite{Umarov10}

\section*{Acknowledgments}
We acknowledge useful conversations with J. P. Gazeau and E. M. F. Curado, and partial financial support from Universidad Polit\'{e}cnica de Madrid, Centro Brasileiro de Pesquisas F\'{\i}sicas, CNPq, Faperj, and Capes (Brazilian agencies).
One of us (GR) acknowledges warm hospitality at CBPF. The other one (CT) acknowledges partial financial support from the John Templeton Foundation.

\end{document}